\documentclass[12pt,preprint]{aastex}

\def\kms{km~s$^{-1}$}
\def\cm2{cm$^{-2}$}

\def\lya{Ly$\alpha$}

\def\mgtwo{\ion{Mg}{2}}
\def\nhi{$N$(H~I)}
\def\cfour{\ion{C}{4}}  

\def\bmk{B$-$K}
\def\bmkn{(B$-$K)$_n$}
\def\ebmv{$E(B-V)$}

\shorttitle{Optical -- Infrared Colors of CORALS QSOs}
\shortauthors{Ellison et al.}

\begin{document}


\title{The Optical -- Infrared Colors of CORALS QSOs:  Searching for Dust
Reddening Associated With High Redshift Damped Lyman Alpha Systems\footnote{These data were 
obtained from the NTT on La Silla (ESO program 072.A-0014(A, B, C, D))}.}


\author{Sara L. Ellison}
\affil{University of Victoria, Dept. Physics \& Astronomy, 
Elliott Building, 3800 Finnerty Rd, Victoria, V8P 1A1, British Columbia,
Canada}
\email{sarae@uvic.ca}

\author{Patrick B. Hall}
\affil{Department of Physics and Astronomy, York University,
4700 Keele St., Toronto, Ontario M3J 1P3, Canada}
\email{phall@yorku.ca}

\author{Paulina Lira}
\affil{Departamento de Astronom\'{\i}a, Universidad de Chile, Casilla 36-D,
Santiago, Chile}
\email{plira@das.uchile.cl}


\begin{abstract}

The presence of dust in quasar absorbers, such as damped
Lyman alpha (DLA) systems, may cause the background QSO to appear
reddened.  We investigate the extent of this potential reddening
by comparing the optical-to-infrared (IR) colors of QSOs with and
without intervening absorbers.  Our QSO sample is based on the
Complete Optical and Radio Absorption Line System (CORALS) survey
of Ellison et al (2001).  The CORALS dataset consists of 66 radio-selected
QSOs at $z_{\rm em} \ge 2.2$ with complete optical identifications.
We have obtained near-simultaneous B and K band magnitudes
for subset of the CORALS sample and supplemented our
observations with further measurements published in the
literature.  In total, we have \bmk\ colors for 42/66 QSOs of which
14 have intervening DLAs.  To account for redshift-related color changes, the
\bmk\ colors are normalized using the Sloan Digital Sky Survey (SDSS) 
QSO composite.  The mean normalized \bmk\ color of the DLA
sub-sample is +0.12, whereas the mean for the no-DLA sample is $-0.10$;
both distributions have RMS scatters $\sim$ 0.5.  Neither a student's
T-test nor a KS test indicate that there is any significant
difference between the two color distributions.
Based on simulations which redden the colors of QSOs with intervening DLAs, 
we determine a reddening limit 
which corresponds to \ebmv\ $< 0.04$ (SMC-like extinction) at 99\% 
confidence (3$\sigma$), assuming that \ebmv\ is the same for all DLAs.  
Finally, we do not find any general correlation between absorber properties
(such as [Fe/Zn] or neutral hydrogen column density) and
\bmk\ color.  The two reddest QSOs with DLAs in our sample have
HI column densities which differ from each other
by an order of magnitude and moderate
gas-to-dust ratios as inferred from chemical abundances. 
One of these two QSOs shows evidence for strong associated absorption 
from X-ray observations, an alternative explanation for its very red color.
We conclude that the presence of intervening galaxies causes
a minimal reddening of the background QSO.

\end{abstract}

%% subject header.  Objects should be in the appropriate "individual"
%% headers (e.g. quasars: individual, stars: individual, etc.) with the
%% additional provision that the total number of headers, including each
%% individual object, not exceed six.  The \objectname{} macro, and its
%% alias \object{}, is used to mark each object.  The macro takes the object
%% name as its primary argument.  This name will appear in the paper
%% and serve as the link's anchor in the electronic edition if the name
%% is recognized by the data centers.  The macro also takes an optional
%% argument in parentheses in cases where the data center identification
%% differs from what is to be printed in the paper.



\section{Introduction}

\subsection{Red Quasars}

Although quasars are usually regarded as objects which are
characterised by an ultra-violet (UV)
excess, the existence of red QSOs has been known for more than
two decades (e.g. Smith \& Spinrad 1980; Bregman et al 1981;
Stein \& Sitko 1984).  
Despite their long history, red QSOs have continued to attract much 
interest in the literature, most notably the on-going debate
to identify the primary cause of their extreme optical--infrared (IR)
colors.  Probably the most popular explanation for red
QSOs is extinction due to dust at the systemic redshift,
a conclusion that is gaining substantial support from
the current generation of large QSO surveys (e.g. Richards et al. 
2003; Hopkins et al. 2004; Glikman et al 2004; White et al. 2003).
There are also convincing examples of individual QSOs
which appear to have been highly reddened due to local (to the
QSO) dust
(e.g. Rawlings et al. 1995; Gregg et al. 2000; Stern et al. 2003).
If dust is indeed a common ingredient in QSOs, the corollary is
that samples based on optically selected sources could potentially miss
the reddened population, whereas radio-selected quasars would
represent a more complete sample (e.g. Webster et al 1995; 
Barkhouse \& Hall 2001).  Indeed, Webster et al. (1995)
have claimed that up to 80\% of QSOs may be hidden from the eyes 
of optical surveys due to dust extinction.

However, some observations do not support the hypothesis of dust
local to the QSO as the cause of reddening.  For
example, millimetre observations (Drinkwater, Wiklind \& Combes 1996) 
of 11 of the reddest Webster
et al. (1995) sources failed to detect CO which argues against the
presence of host galaxy dust in these cases. Other
possible sources of QSO `reddening' include the effect of host
galaxy starlight and synchrotron radiation (Francis, Whiting \& Webster 
2000; Whiting, Webster \& Francis, 2001; Benn et al. 1998).  

\subsection{Dust Reddening Associated with Intervening Galaxies}

One factor that could have a significant impact on reddening QSO light
that remains relatively unexplored is the presence of
intervening dust located in foreground galaxies.  The identification
of a handful of red lensed QSOs lends credibility to this
scenario, in which dust in the lensing galaxy reddens the background
source (e.g. Gregg et al. 2000; Falco et al. 1999, Wucknitz et al 2003).
The issue of dusty intervening galaxies is of interest not only
in order to assess its contribution to the red QSO population, but is also of
crucial importance for the study of quasar absorption lines.  Quasar
absorbers, such as the damped Lyman alpha (DLA) systems 
which correspond to the signatures of high redshift galaxies
imprinted on the continuum of a background QSO, represent one of
our best tools for studying the early galaxy population.  Previous
surveys for DLAs have almost universally relied on large optical
databases of QSOs to find intervening absorbers.  Since only 
relatively bright, optically
selected QSOs have been used to identify DLAs, there has been
widespread concern that dusty intervening galaxies 
may obscure the
background QSO so that it drops below the survey magnitude limit.
Such a selection effect would result in a biased view of high 
redshift galaxies.  This general concern has been quantified
in a number of theoretical works which have attempted
to estimate the amount of reddening that a QSO may suffer
and consequently what fraction of the DLA population, and their
contribution towards the cosmic gas and metals budget, may
have been overlooked (e.g. Fall \& Pei 1993; Pei \& Fall 1995;
Pei, Fall \& Hauser 1998; Churches et al. 2004;
Vladilo \& P\'{e}roux, 2005).  However, until recently there
had been little observational progress in quantifying
the number of missed DLAs due to extinction effects.

The first systematic program to quantify dust bias was
undertaken by Ellison et al. (2001) by searching for DLAs in an
optically complete radio-selected 
sample of QSOs.  The Complete Optical and Radio Absorption Line
System (CORALS) survey was
based on a homogeneous sample of 66 radio-selected, flat spectrum 
QSOs taken from
the Parkes catalogue, complete down to 0.25 Jy.  Extensive
optical follow-up observations resulted in identifications
for \textit{all} sources.  Therefore, subsequent
spectroscopic searches for DLAs were based on a complete sample,
ideal for determining the extent to which intervening galaxies
obscure background QSOs. Ellison et al. (2001) determined that the neutral 
gas mass density of DLAs ($\Omega_{\rm DLA}$)
may have been previously under-estimated by at most a factor of
$\sim$ 2.  However, this work also found tentative evidence
that $\Omega_{\rm DLA}$ and $n(z)$ (the DLA number density)
may be under-estimated for 
bright QSO samples with magnitude limits $B \lesssim 19.5$.  Ellison et 
al. (2004) re-visited this issue and suggested that this could be
understood in terms of an Eddington bias if the optical luminosity
function deviates from a simple power law.
%This conclusion has recently been strengthened
%by a similar survey which has doubled the number the redshift
%path of CORALS (REF).
To extend the original CORALS sample to lower redshifts, Ellison et al. (2004)
studied \mgtwo\ absorption systems in an analogous radio-selected QSO sample.
The main conclusion of that work was that the number density 
of intervening systems selected 
via \mgtwo\ absorption in the range $0.6 < z < 1.7$
is in excellent agreement with magnitude
limited samples (e.g. Nestor, Turnshek \& Rao 2005).  Most recently, Akerman
et al. (2005) have measured the metal abundances of $z>1.8$ CORALS
DLAs and found that the weighted mean metallicity based
on zinc (an undepleted proxy for Fe) is only marginally higher
(but consistent within the error bars) than previous samples.

Although the results from CORALS indicate that magnitude
limited surveys (at least those complete to $B \sim 19$)
have not missed a significant fraction of the gas or
metals budget, this is not to say that DLAs are dust-free.  In fact,
it has been known for over a decade, based on the ratios of refractory
and non-refractory elements, that dust is present in DLAs (Meyer \&
Roth 1990; Meyer \& York 1992; Pettini et al. 1994, 1997).  
Understanding the level of extinction caused by this dust has so far yielded 
conflicting results.  For example, Fall, Pei \& McMahon (1989)
and Pei, Fall \& Bechtold (1991)
found a significant difference between the spectral indices of QSOs 
with intervening DLAs and a control sample.  However, this result has not
been reproduced with larger, homogeneous QSO samples.  Currently, the
most stringent limit for reddening due to intervening DLAs has been
obtained from fitting spectral indices to $\sim$ 1450 QSOs from the 
Sloan Digital Sky Survey (SDSS): \ebmv\ $<$ 0.02 for SMC-like extinction 
(Murphy \& Liske 2004).  Despite this sensitive limit, studies of spectral 
indices suffer from two drawbacks:
they utilize limited wavelength coverage in the optical regime and have not
yet been 
conducted on unbiased samples (i.e. they were taken from magnitude
limited surveys).  

\subsection{Optical--Infrared Colors and Extinction Curves}

In this work, we take a different approach to studying the effect
of dust extinction from intervening DLAs by using optical--IR
QSO colors to search for a reddening signature.  This technique has 
the advantage of using a wide wavelength baseline, from the
observed frame B to K bands, over which to detect 
the effect of extinction.  Moreover, by studying QSOs from the CORALS survey,
we have a sample which is optically complete and in which all of the 
intervening high redshift DLAs have already been identified and
metallicities measured.

\begin{figure}
\centerline{\rotatebox{270}{\resizebox{13cm}{!}
{\includegraphics{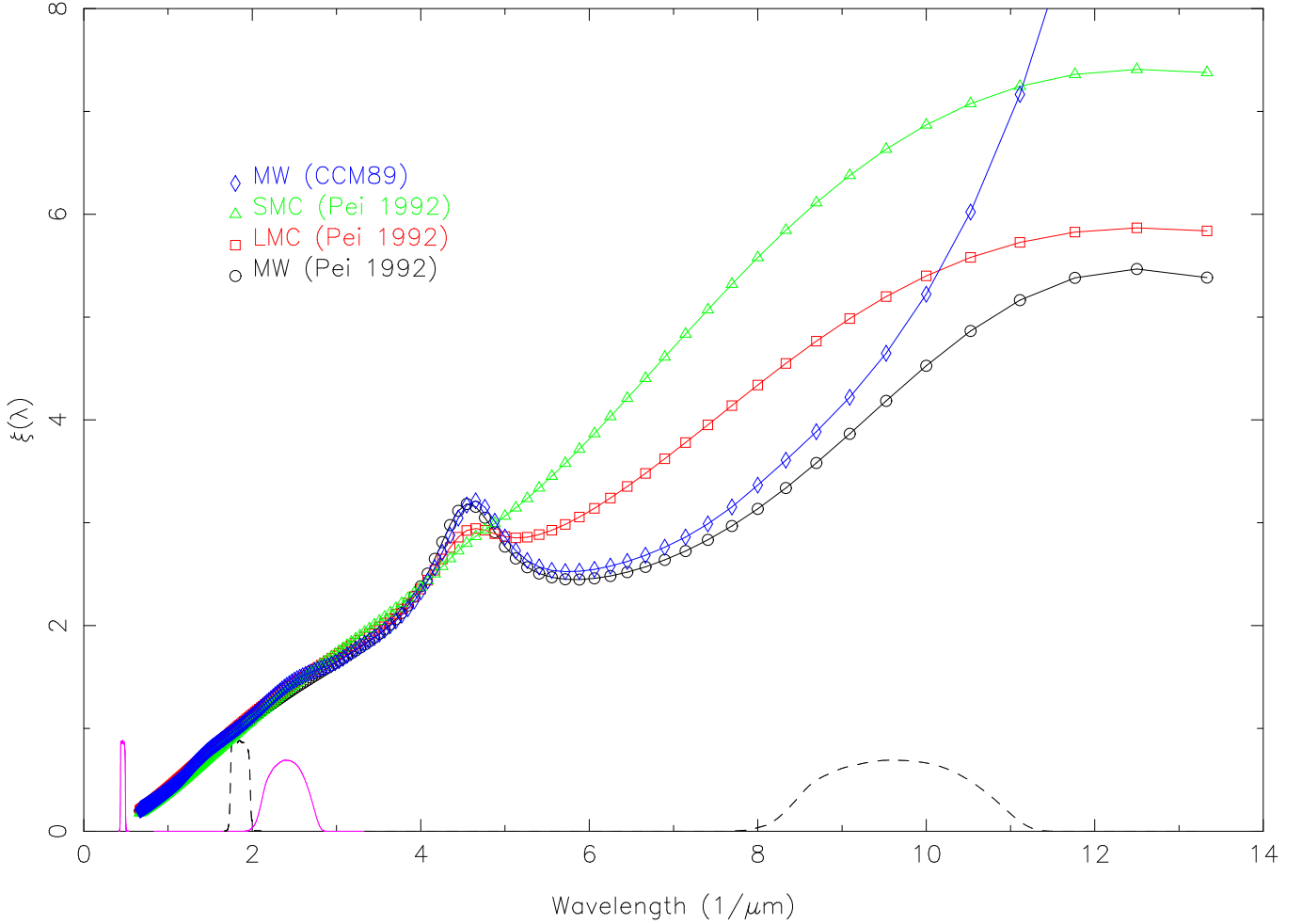}}}}
\caption{\label{ext_curve} Extinction curves for the parameterizations
of Pei (1992) and Cardelli, Clayton \& Mathis (1989).  Assumed
$R_V$ values are 2.93 (SMC), 3.16 (LMC) and 3.08 (MW). There were no
data available at wavelengths $\le 1000$ \AA, hence the divergence
in the FUV for the Galactic curves represents an extrapolation of
the fitting functions.  Filter transmission curves along the base
of the plot illustrate what rest wavelengths correspond to the B
and K$_s$ bands for $z=0$ (solid) and $z=3$ (dashed).}
\end{figure}

The observed reddening obviously depends
largely on the assumed extinction law, although the QSO and
galaxy (i.e. dust) redshift also play a part (as we demonstrate below).  
Despite its
ubiquity in astrophysical environments, the properties of
dust remain relatively poorly understood (see reviews by Mathis 1990
and Draine 2003) and reliable
empirical extinction data are available for only the Milky Way (MW) and
Magellanic Clouds.  In particular, data are sparse in the
far-ultraviolet (FUV, Hutchings \& Giasson 2001) and below the 
Lyman limit there is simply
no flux with which to work.  FUV wavelengths are particularly important
for the present work, which considers the optical--IR colors of
high redshift QSOs.  The most often used parameterization of Galactic
extinction is that of Cardelli, Clayton \& Mathis (1989; hereafter
CCM89).  The advantage of the CCM89 formulae is that $\xi(\lambda) = A_{\lambda}/
A_V$ can be calculated for any value of $R_V$\footnote{In the usual
notation, $R_V = A_V /$\ebmv, i.e. the total-to-selective
extinction.}. Another popular
parameterization of Galactic extinction is that of Pei (1992),
although this is for a fixed value of $R_V = 3.08$.  A comparison
of these two extinction law fits is shown in Figure \ref{ext_curve},
with an $R_V$ = 3.08 for the CCM89 law in order to match the Pei (1992) curve.
It can be seen that at IR and optical wavelengths, the two
curves are in close agreement.  However, in the FUV ($\lambda < 1200$ \AA)
the curves diverge; this is purely due to extrapolation of
the different fitting functions into a regime where no data were available.  
Whereas an extrapolation of the
Pei (1992) parameterization leads to a flattening of $\xi(\lambda)$,
the CCM89 curve rises quickly at FUV wavelengths.  In Figure \ref{ext_curve}
we also show the extinction curves of the SMC and LMC as fitted
by Pei (1992), again extrapolated to the FUV using the published
fitting functions.

\begin{figure}
\centerline{\rotatebox{270}{\resizebox{13cm}{!}
{\includegraphics{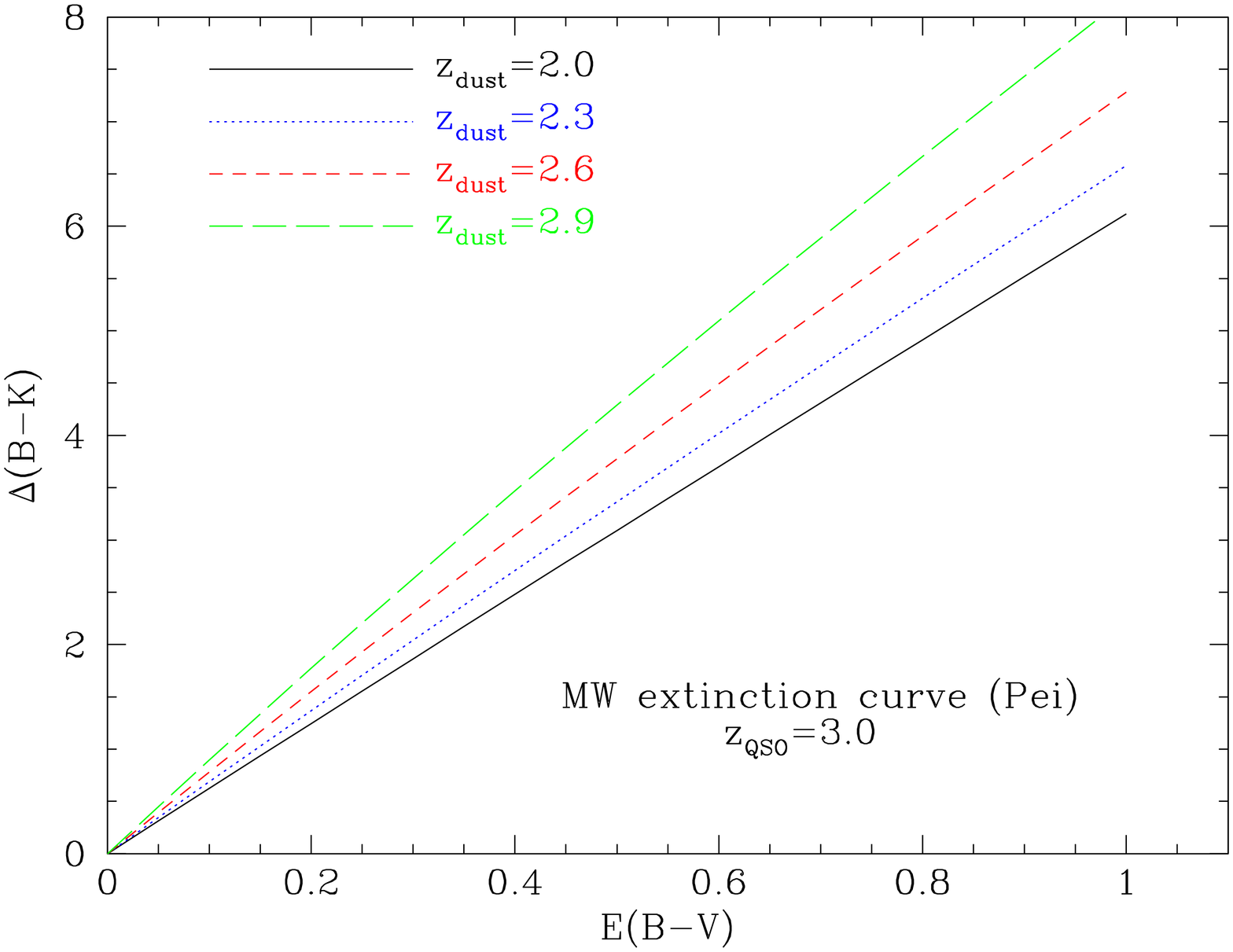}}}}
\caption{\label{BmK_theo_MW_Pei} Theoretical reddening of a $z_{\rm em}=3.0$
QSO with an intervening absorber at $z_{\rm abs} = 2.0, 2.3, 2.6, 2.9$.
We assume a Galactic extinction curve based on the parameterization
of Pei (1992). }
\end{figure}

\begin{figure}
\centerline{\rotatebox{270}{\resizebox{13cm}{!}
{\includegraphics{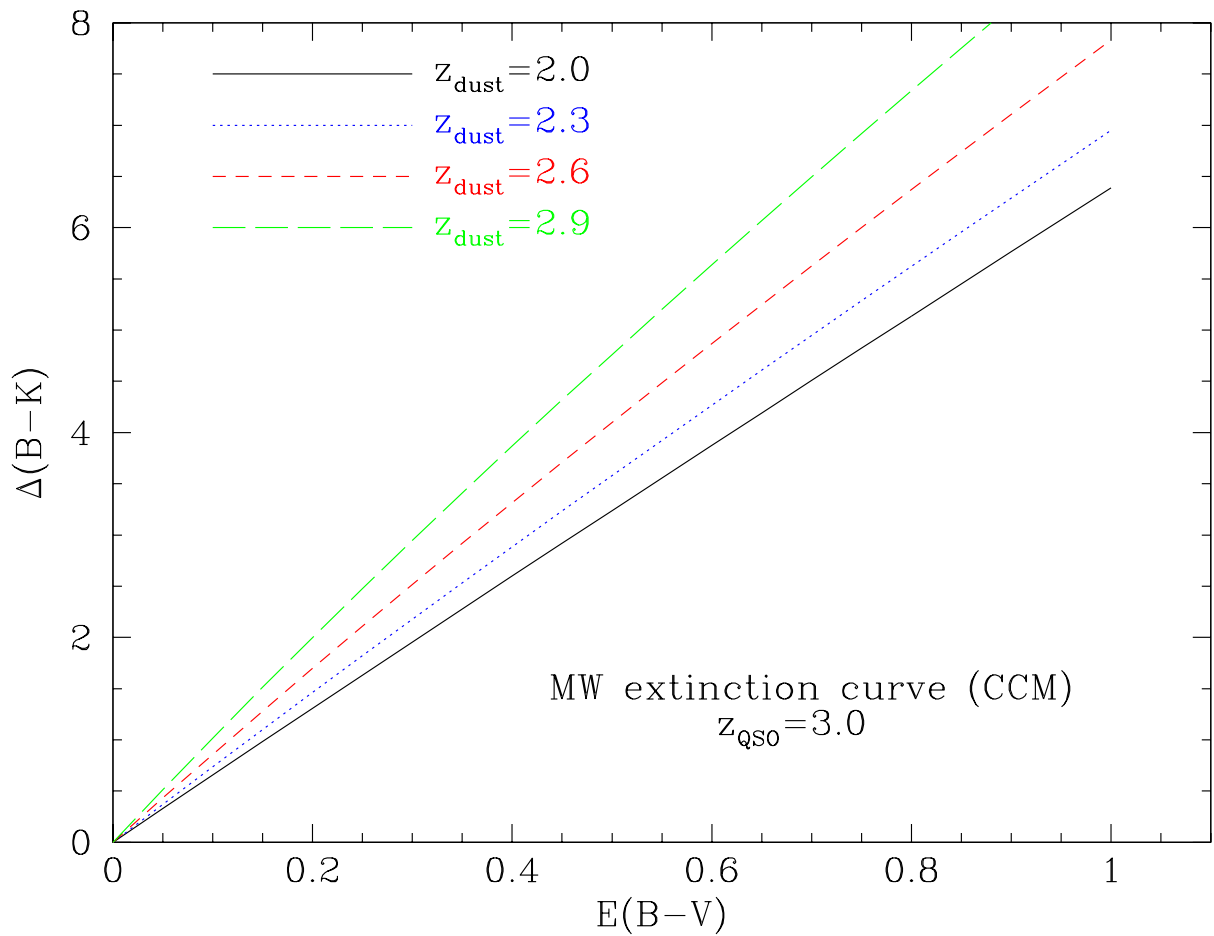}}}}
\caption{\label{BmK_theo_MW_CCM} Theoretical reddening of a $z_{\rm em}=3.0$
QSO with an intervening absorber at $z_{\rm abs} = 2.0, 2.3, 2.6, 2.9$.
We assume a Galactic extinction curve based on the parameterization
of Cardelli, Clayton \& Mathis (1989) with $R_V = 3.08$. }
\end{figure}

\begin{figure}
\centerline{\rotatebox{270}{\resizebox{13cm}{!}
{\includegraphics{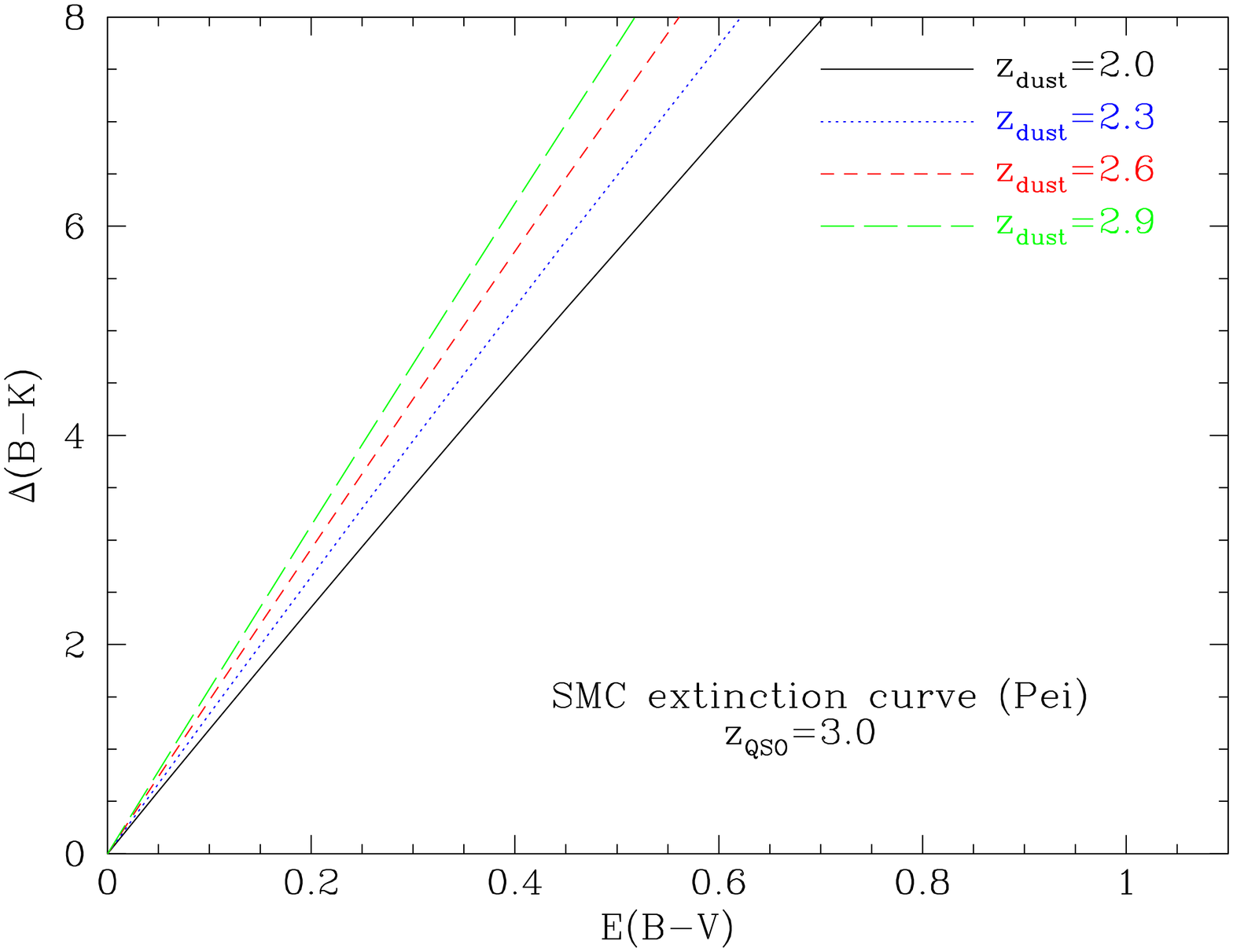}}}}
\caption{\label{BmK_theo_SMC_Pei} Theoretical reddening of a $z_{\rm em}=3.0$
QSO with an intervening absorber at $z_{\rm abs} = 2.0, 2.3, 2.6, 2.9$.
We assume an SMC extinction curve based on the parameterization
of Pei (1992). }
\end{figure}

We illustrate the expected reddening due to intervening dust in   
Figures \ref{BmK_theo_MW_Pei}, \ref{BmK_theo_MW_CCM} and 
\ref{BmK_theo_SMC_Pei}.
In each figure, we plot the theoretical reddening (as measured by a 
change in the \bmk\ color) of a QSO at $z_{\rm em} = 3$ as a function 
of \ebmv.  Since the color change depends on the redshift of the
intervening dust, four different lines are plotted for dust at
$z_{\rm dust} = 2.0, 2.3, 2.6, 2.9$.  The reddening induced by the
two Galactic extinction curves is quite similar, although a CCM89
law produces slightly redder colors due to the rapid rise of $\xi(\lambda)$
in the FUV.  Reddening is most severe for SMC type dust, and could
be even more extreme if the true extinction law continues to
rise at FUV wavelengths. That is, the parameterization of SMC dust
by Pei (1992) may be considered conservative.

\section{Observations and Data Reduction}

Since flat spectrum radio-loud QSOs (RLQs) are highly variable, 
it is important to obtain
optical and IR data from near-simultaneous epochs 
(e.g. Francis, Whiting \& Webster 2000, hereafter FWW00).  We achieved
this by using the optical (SuSI2) and IR (SofI) imagers on the New
Technology Telescope (NTT) at La Silla observatory.  Since both 
instruments are mounted on the same
Nasmyth platform, the overhead incurred by switching between
them is typically only about 2 minutes, depending on the source
position relative to the zenith.  Our strategy was therefore to
observe a sequence of QSOs in the optical and then change to
the IR.  Optical and IR photometric standard stars were observed 
throughout the night at airmasses up to 1.8.

Our data were obtained during two runs in October 2003 and March 2004.
The March 2004 run experienced excellent conditions: photometric
transparency and typical seeing $\sim$ 0.6 arcsec.  The earlier
October 2003 run, suffered from somewhat poorer seeing ($\sim$ 1 arcsec)
and the end of the last night (during which only SofI was used)
was not photometric.

\subsection{Optical Imaging}

In addition to B-band images, we obtained supplemental V-band images for
most QSOs with redshifts $z_{\rm em} > 3.0$.  
Typical exposure times in the B and V bands ranged from 10 to 200
seconds (see Table \ref{obs_table}), the total integration was
divided between two
exposures with a spatial offset of a few hundred pixels between exposures.
We used SuSI2 in 2x2 binning mode which yielded a pixel scale of 0.16 
arcsec/pixel.  Standard IRAF procedures were used for the reduction
of the optical data:  a bias was constructed from a median combination
of five frames and the overscan was fitted and subtracted interactively.
The initial flat-fielding was executed using dome-flats to correct
pixel-to-pixel variations, followed by an illumination correction
of large-scale variations constructed from a super-sky flat.
Instrumental magnitudes were determined for each night
using IRAF's \textit{qphot} 
package with between 20 and 40 standard stars taken from Landolt (1992).
The photometric solution (zero points and extinction coefficients)
for each night in the B and V bands were determined by a least squares
fit to the data.  The zero points determined in this way indicate that 
conditions were
photometric throughout the optical observations.  The RMS error
in the photometric solution was 0.04 magnitudes in the V and B bands
for the October 2003 run and 0.04 and 0.07 for V and B bands respectively
in March 2004.  The final
quoted $B$ and $V$-band errors in Table \ref{mag_table} include the 
uncertainty in the fit of the photometric
solution and photon statistics added in quadrature.

\subsection{Infrared Imaging}

Each SofI K$_s$ band integration consisted of a series of co-added 10 second
exposures with a random offset within a 40 arcsec jitter box between
each exposure.  Total on-source integration times ranged between 100 and
5400 seconds (see Table \ref{obs_table}).  IR data reduction was carried out 
using Pat Hall's
Infrared Imaging Reduction Software (PHIIRS) which incorporates
many standard IRAF routines.  We did not follow the usual procedure of 
obtaining dark frames due to the non-linearity of dark current with incident
flux.  Instead, we used the PHIIRS task {\em irsky} to combine individual 
dithered exposures to simultaneously remove dark current and sky.
We used 4--6 frames to construct the sky image for each science
exposure; fewer frames were used when the telescope was observing
at low airmass, since pupil ghosts produce more severe artifacts
when the pupil plane rotates more quickly.  These ghosts
remained in a few of the final science images, but do not affect
the central part of the array where the QSO is located.
Flat fielding was done using SofI's `special dome flats' and custom
written reduction script available from the NTT web pages.
Registration of the dithered images and median co-addition
were done with the PHIIRS routines
\textit{irshift} and \textit{ircoadd}, scaling
the images to the mode of the sequence.  Bright objects were
identified in the co-added frame and masked out before running
through the co-addition process a second time.

The photometric solution was determined in a manner similar to
the optical observations using IR photometric standards
from Persson et al. (1998).  Although the March 2004 run was photometric,
cirrus was present at the end of the October 2003 run.  For the
affected SofI data, we determined K$_s$ magnitudes by boot-strapping
the photometric solution from 2MASS point sources. Typically five
2MASS point sources were available per field for the bootstrap
calibration, the larger errors for the magnitudes of these QSOs
reflect the RMS in the fit to 2MASS magnitudes\footnote{One QSO,
B0537$-$286, was observed during both the October 2003 and
March 2004.  The magnitude determined from the bootstrap analysis,
$K = 16.11 \pm 0.04$ agrees (within the errors) with the magnitude 
determined under photometric conditions.}
A filter transformation\footnote{http://www.ls.eso.org/lasilla/sciops/ntt/sofi/setup/Zero\_Point.html} was applied
in order to place the SofI magnitudes on the Persson $K_s$ system.  However,
in the absence of contemporaneous J-band magnitudes
we have estimated the J$-$K color as a function of
\bmk\ using the dataset of
FWW00.  Given the scatter in this color-color relation, we
expect that the error in assuming a J$-$K color will affect the
filter transformation at $<0.02$ magnitudes.  The final
quoted $K_s$-band errors in Table \ref{mag_table} include the uncertainty 
in the fit of the photometric
solution, photon statistics and the estimated error in the filter
transformation added in quadrature.

Finally, both optical and IR colors were corrected for Galactic
extinction according the the maps of Schlegel, Finkbeiner \&
Davis (1998).
Our final optical and IR magnitudes are presented in  Table \ref{mag_table}.

\subsection{Literature Magnitudes}

We have supplemented our measurements in Table \ref{mag_table} with magnitudes
determined by FWW00, who also achieved near-simultaneous optical and IR 
($<$ 6 nights separation) observations of a number of CORALS QSOs.
In Table \ref{mag_table} we have corrected the K$_n$ band magnitudes (deduced
from the SAAO system) given 
in FWW00 to K$_s$ magnitudes by using the conversion
determined by 2MASS\footnote{http://www.astro.caltech.edu/~jmc/2mass/v3/transformations/}.  We also corrected the photometry of FWW00 for
Galactic extinction according to the maps of Schlegel et al (1998).

With these supplemental data, we have optical--IR colors for 45/66
program QSOs, with \bmk\ colors for 42.  Of these 42, 14 have intervening 
DLAs; only 3 QSOs
with four intervening DLAs were not observed by us or FWW00\footnote{
Ellison et al. (2001) reported 19 intervening DLAs.  However,
one of these, the $z_{\rm abs} = 1.875$ DLA towards B2314$-$409
has been shown by Ellison
\& Lopez (2001) to have an \nhi\ slightly below the canonical
DLA limit of $2 \times 10^{20}$ \cm2.}.
The observed sub-sample of CORALS QSOs are representative of
the whole sample; only scheduling restrictions prevented the
observation of the full sample.  
In the remainder of this paper we refer the colors determined
from our B and K$_s$ magnitudes as \bmk.

\section{Results}

\subsection{Comparison of Optical and IR Colors}

\begin{figure}
\centerline{\rotatebox{270}{\resizebox{13cm}{!}
{\includegraphics{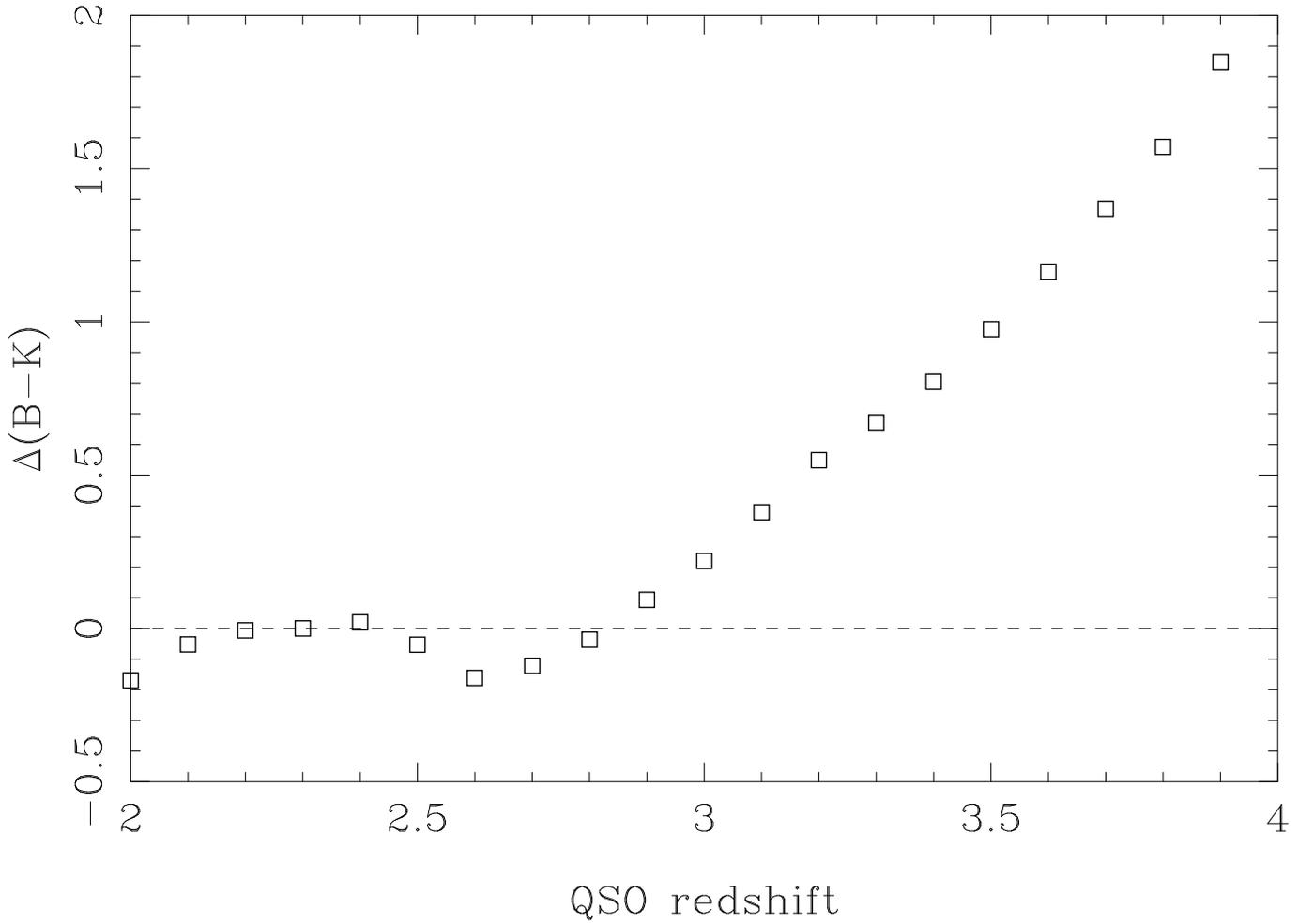}}}}
\caption{\label{bmk_template} Change in \bmk\ color as a function of redshift
determined using the SDSS composite spectrum of Vanden Berk (2001).
Color changes are relative to $z=2.3$.}
\end{figure}

The observed \bmk\ color of a given QSO will change with redshift
as the intrinsic spectrum slides through the filter set.  Strong
emission features such as \lya\ and \cfour\ can increase the
observed flux in a given band, although flux depression due
to the \lya\ forest is the dominant source of color shift 
(e.g. Richards et al. 2003).  In Figure \ref{bmk_template}
we demonstrate this effect by calculating the change in \bmk\
as a function of redshift using the SDSS composite spectrum
of Vanden Berk (2001).  The color is calculated relative to
a reference redshift, in this case $z=2.3$.  We therefore
determine a normalized \bmk, \bmkn, which is calculated in the
following way.  First, we determine the median \bmk=2.92 for QSOs in the
range $2.2 < z < 2.4$ since the color changes very little over this
interval (see Figure \ref{bmk_template}).  Next, we determine
a redshift dependent color correction based on the SDSS composite
spectrum, displayed graphically in Figure \ref{bmk_template}.
In the case of QSOs with intervening DLAs, we make
a correction ($\Delta  (B-K)_{DLA}$) for the suppression of flux 
in the $B-$band
due to the \lya\ line.  For each DLA, we simulate a spectrum
with a damped profile at the redshift $z_{\rm abs}$ and with \nhi\
given in Table \ref{norm_bmk_tab}.  The spectrum is multiplied by the filter
transimission curve and the resulting magnitude compared with
an identical spectrum without a DLA absorber.  Typical corrections
are a few hundredths to a tenth of a magnitude, with a 
maximum $\Delta  (B-K)_{DLA} = 0.18$.
The final, normalized \bmk\ color of each QSO is calculated by subtracting
the median, the redshift dependent term and (in the case of an intervening
DLA) the absorption correction term 
[$(B-K)_n=(B-K)-2.92-f(z)-\Delta  (B-K)_{DLA}$]. 

\begin{figure}
\centerline{\rotatebox{270}{\resizebox{13cm}{!}
{\includegraphics{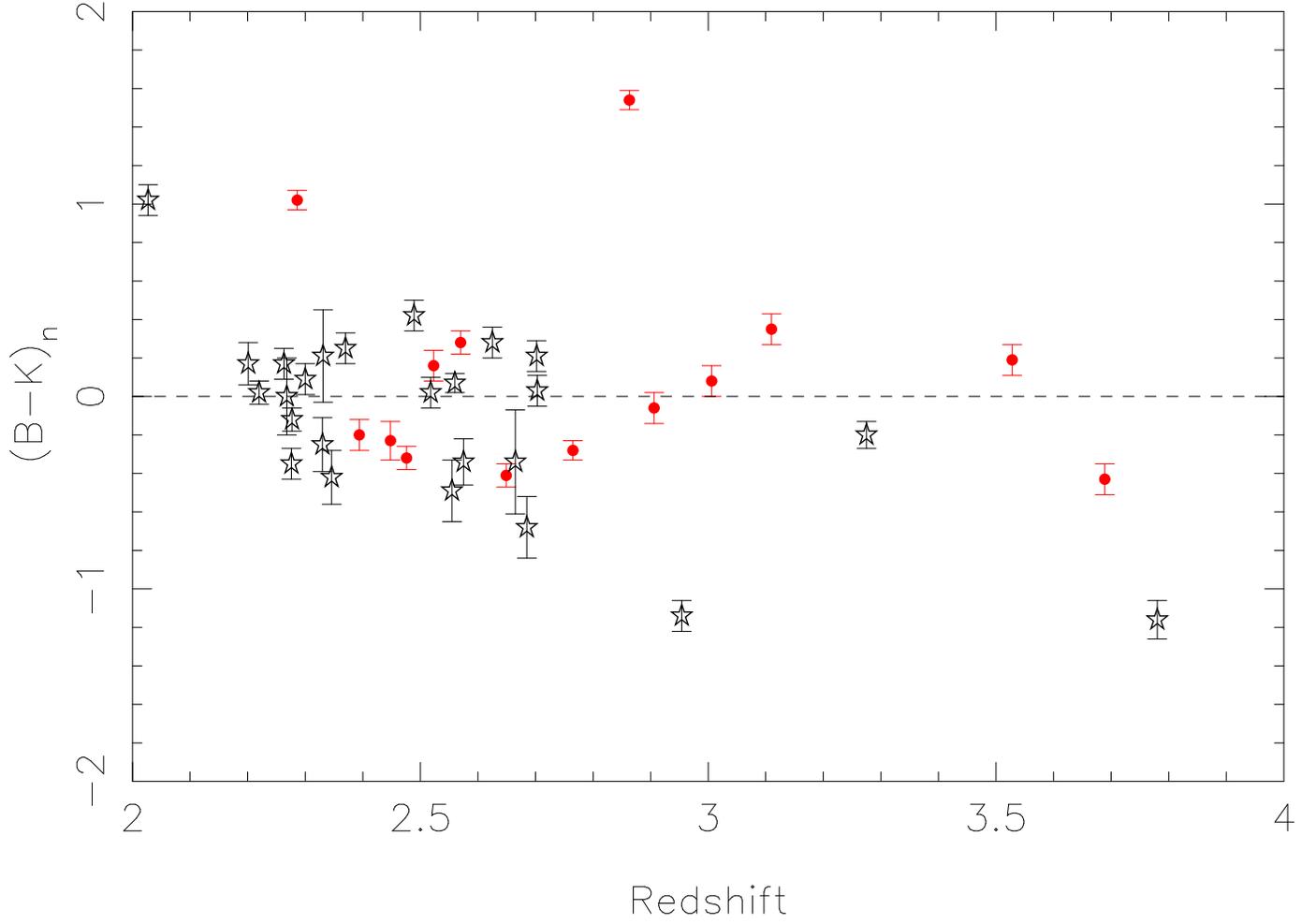}}}}
\caption{\label{norm_z} Redshift distribution of normalized \bmk\ colors for QSOs
with (filled circles) and without (open stars) DLAs.}
\end{figure}

Table \ref{norm_bmk_tab} presents the normalized \bmk\ colors
for the procedure described above and 
Figure \ref{norm_z}\footnote{In this Figure, Table \ref{norm_bmk_tab} and in
the following statistics, we only include QSOs whose $B$ and $K_s$
band magnitudes have errors less than 0.3 magnitudes.  In practice,
this excludes only 3 of the FWW00 QSOs which have poorer photometry.
Our conclusions are unaffected by the exclusion of these three.} 
demonstrates that the resulting color distribution
does not exhibit any redshift dependence.  There is, however,
a large spread in \bmkn\ at any given redshift, regardless of whether 
or not a DLA is
present, as expected for RLQs (e.g. FWW00).    

\subsection{Limits on Extinction}

In order to determine quantatatively whether the colors of QSOs with
and without DLAs are drawn from the same distribution, we execute a 
Kolmogorov-Smirnov (KS) test on two sub-samples: the 14 QSOs
with DLAs and 25 without. The KS test yields a 94\% probability that 
the two sub-samples are drawn from the same distribution.  The mean 
(median) normalized
colors of QSOs with DLAs are +0.12 (+0.01) and $-0.10$ ($+0.02$) for
QSOs without DLAs with RMS scatters of 0.56 and 0.47 respectively.    
A student's T-test yields a
test statistic = 1.32, indicating that the mean colors of
the two samples are not significantly different.  

In order to quantify a limit to the amount of reddening associated
with the DLAs in our sample, we run a series of 
simulations that artificially include/remove reddening in the spectra
of QSOs with DLAs.  We then 
calculate the KS statistic compared with the no DLA
sample. In each simulation, we redden the observed \bmk\ color of those QSOs
with DLAs assuming a fixed \ebmv\ for each DLA for both Galactic and
SMC extinction curves.  
We use the parameterizations of SMC  ($R_V= 2.93$)
and Galactic ($R_V = 3.08$) extinction from
Pei (1992) and a second functional fit to Galactic reddening from
CCM89, as described in section 1.3.    
The colors of QSOs without DLAs remain unchanged in these simulations.

%We also assume one of two
%recipes for allocating an \ebmv\ to each DLA; either we fix
%\ebmv\ for every DLA in a given run, or we assign a value which correlates
%with the \nhi.  Such reddening-to-gas\footnote{Some authors refer to the
%relationship between \ebmv\ and \nhi\ as a dust-to-gas ratio.  However,
%we prefer to use the term reddening-to-gas ratio to avoid confusion
%with the parameters $k$ and $\kappa$ that are often used in the literature
%to denote dust-to-gas ratios inferred, for example, from chemical
%abundances.}  

\begin{figure}
\centerline{\rotatebox{270}{\resizebox{13cm}{!}
{\includegraphics{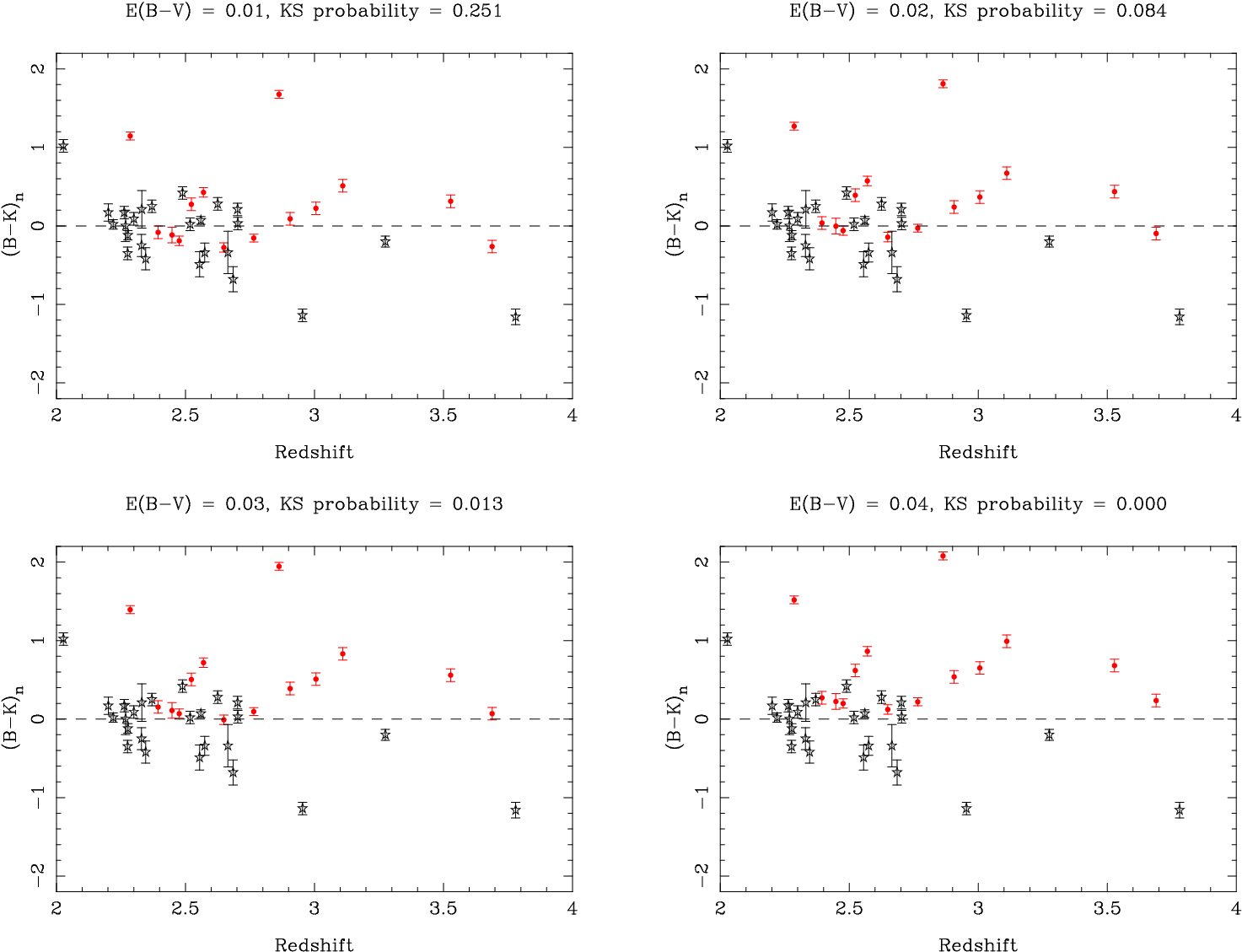}}}}
\caption{\label{ks_fig_add} Redshift distribution of normalized \bmk\ colors 
for QSOs with (filled circles) and without (open stars) DLAs with
reddening \textit{added} to the former population assuming a Pei (1992) SMC
extinction law.  Each of the four panels illustrates a different degree
of reddening and shows the KS probability that the DLA and non-DLA
QSOs are drawn from the same color distribution.}
\end{figure}

\begin{figure}
\centerline{\rotatebox{270}{\resizebox{13cm}{!}
{\includegraphics{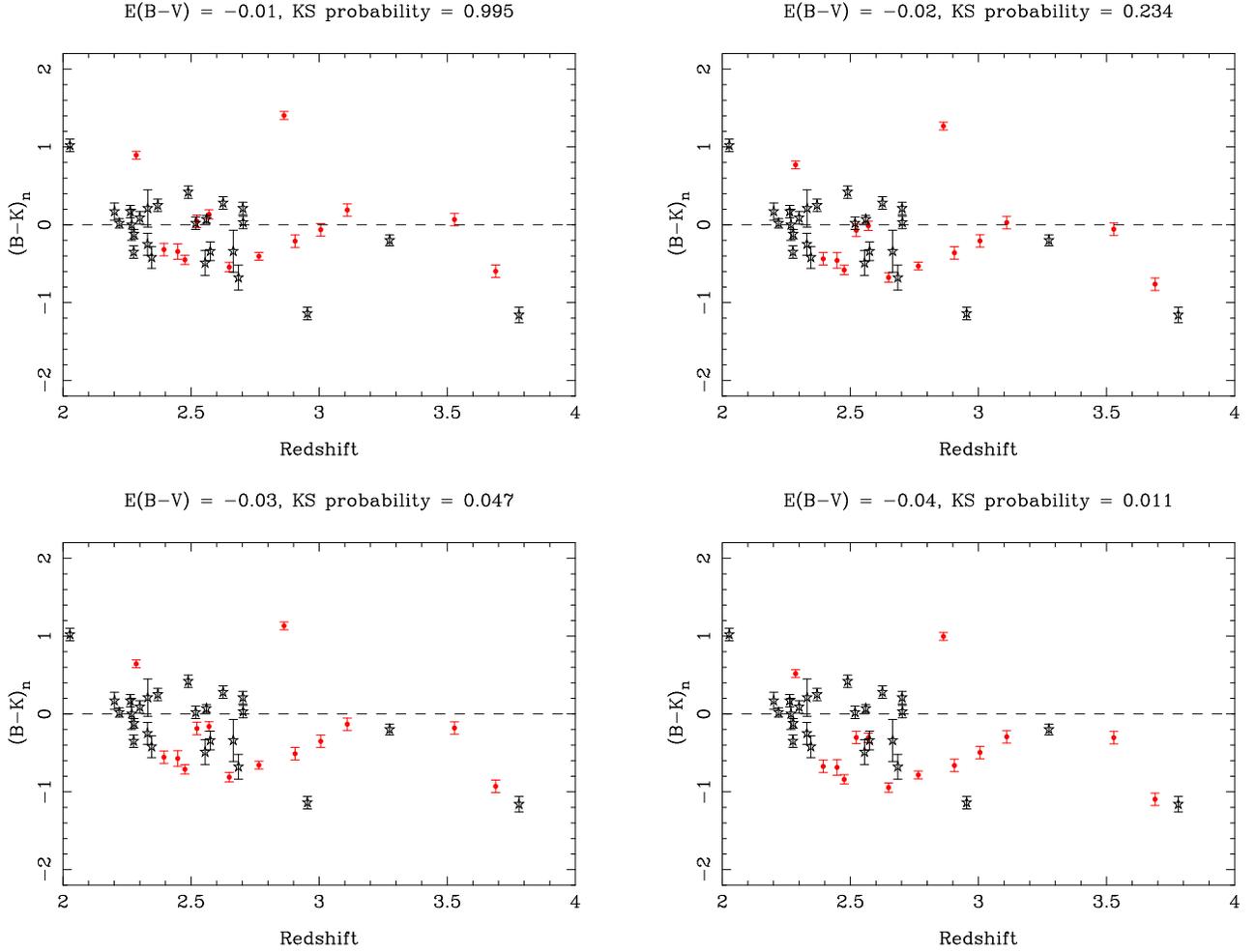}}}}
\caption{\label{ks_fig_remove} Redshift distribution of normalized \bmk\ colors 
for QSOs with (filled circles) and without (open stars) DLAs with
reddening \textit{removed} to the former population assuming a Pei (1992) SMC
extinction law.  Each of the four panels illustrates a different degree
of reddening and shows the KS probability that the DLA and non-DLA
QSOs are drawn from the same color distribution.}
\end{figure}

\begin{figure}
\centerline{\rotatebox{270}{\resizebox{13cm}{!}
{\includegraphics{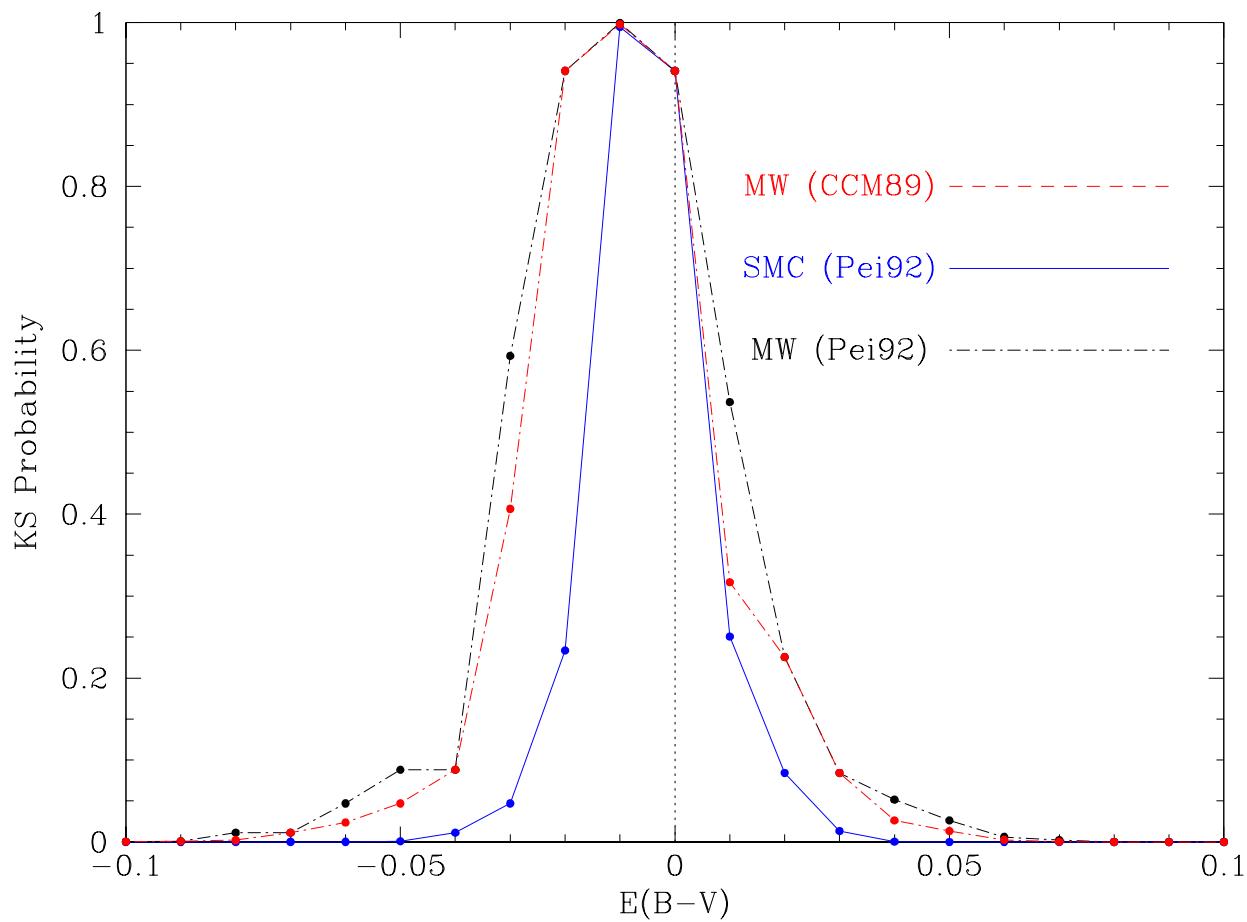}}}}
\caption{\label{ks_all} KS probability that the \bmkn\ colors of 
QSOs with and without DLAs are drawn from the same distribution, for
simulations which include reddening in the former population.  Positive
values of \ebmv\ indicate that the effect of dust (i.e. reddening) was
added to the simulated spectra and negative values indicate that
the effect of dust was removed.  The vertical dotted line indicates
that the raw \bmkn\ have been used, i.e. no reddening included.}
\end{figure}

To demonstrate
how a small change in \ebmv\ can affect the color distribution and
KS statistic we show the \bmk\ colors for a range of simulations 
in Figures \ref{ks_fig_add} and \ref{ks_fig_remove}.  The corresponding
KS probability is given above each panel.  The results of the full
simulation gamut for the 
three extinction curves are shown in Figure \ref{ks_all}, where a positive
\ebmv\ indicates that the spectrum was reddened, whereas a negative value
indicates that the effects of dust were removed.  These simulations
rule out \ebmv $>0.04$ (SMC) and \ebmv $>0.07$ (MW) at 99\%
confidence (i.e. KS probability $\le 0.01$ for \ebmv\ $<-0.04, -0.07$,
with no difference between the Pei and CCM89 Galactic models).

Alternatively, we can consider a reddening that depends on the
\nhi\ of each DLA.  Reddening-to-gas\footnote{Some authors refer to the
relationship between \ebmv\ and \nhi\ as a dust-to-gas ratio.  However,
we prefer to use the term reddening-to-gas ratio to avoid confusion
with the parameters $k$ and $\kappa$ that are often used in the literature
to denote dust-to-gas ratios inferred, for example, from chemical
abundances.} ratios have been determined in local galaxies, e.g.

\begin{equation}\label{smc_eqn}
E(B-V) = \frac{N(HI)}{4.0 \times 10^{22}}
\end{equation}

for the SMC (Bouchet et al. 1985) and 

\begin{equation}\label{mw_eqn}
E(B-V) =\frac{N(HI)}{5.8 \times 10^{21}}
\end{equation}

for the Milky Way (Bohlin, Savage \& Drake 1978).  

If reddening is removed from the DLA-QSO sample in accordance with
the SMC reddening-to-gas relation in Equation \ref{smc_eqn} the KS 
probability is 0.39, an inconclusive result.  
Correcting the colors of QSOs with DLAs using the more extreme
Galactic relation
in Equation \ref{mw_eqn} results in colors that are significantly
bluer than the no-DLA QSOs, i.e. the reddening correction is
too large.  The KS probability in this case is 0.03, ruling
out a Galactic reddening-to-gas relation at 2 $\sigma$.

In reality, reddening is likely to be more stochastic than the 
simple dust recipes adopted here, and extinction 
curves are likely to vary from DLA to DLA.    
Our dataset is currently too small to quantify the reddening-to-gas ratios 
for DLAs through modelling techniques such as $\chi^2$ minimization, 
although we can rule out a Galactic scaling relation
at 97\% confidence.   However, future work with larger samples,
e.g. with DLAs identified from the SDSS, will be able to investigate
the reddening-to-gas relation in more detail.  
Finally, we note that a detailed treatment of other such effects as
non-negligible intergalactic medium reddening, host galaxy
contributions or exotic extinction laws are beyond the scope of
this paper.

\subsection{X-ray Observations of the Reddest QSOs}

To further investigate the 3 QSOs with
the reddest \bmk\ colours, we have searched for archival X-ray
observations of these sources. Given a sufficient signal-to-noise in
the X-ray data, it is possible to investigate whether `intrinsic'
absorption (i.e., associated with the QSO) is present. In principle 
the X-ray data can measure, or place constraints upon, the integrated 
hydrogen column along the line of sight (and by assuming a 
dust-to-gas ratio, on the reddening), 
as well as the redshift at which the absorption takes place.  
Since the inferred \nhi\ column
densities of intrinsic absorbers are frequently an order of magnitude
higher than typical intervening DLAs, it is plausible that significant 
reddening may be induced by such an absorber (although an important 
caveat is the large possible range in the reddening-to-dust 
ratios).  

\medskip

Of the three sources with \bmkn\ $ > 1$ we found that X-ray
observations had been obtained for the QSOs B0438$-$436 and
B0458$-$020, both of which exhbit DLA absorption. 
Unfortunately, no X-ray observations have been obtained
for B1318$-$263, the system without a DLA absorber.

\medskip

B0438$-$436 has long been known to exhibit significant excess (i.e.
above the Galactic contribution) absorption. 
Evidence for this excess absorption has
come from the analysis of ROSAT, ASCA and XMM observations
(Serlemitsos et al. 1994; Elvis et al. 1994; Cappi et al. 1997; Brocksopp
et al 2004). The best quality data so far are those obtained with XMM
and the best-fit model (Brocksopp et al 2004) is consistent with a
solar metallicity
absorber with N(H) $= 1.3 \times 10^{22}$ cm$^{-2}$, which Brocksopp 
et al. assumed to be at the redshift of the QSO\footnote{With sufficiently
high quality data, it is possible to determine the absorption redshift.
However, associated absorption is often the working assumption since
the column densities greatly exceed that of typical intervening absorbers.
In this case, the intervening DLAs has \nhi\ $= 6 \times 10^{20}$ \cm2.}.
If we assume that ionized and molecular gas is negligible,
this column density translates into a $E(B-K) \sim 18.5$ and $\sim
5$, for a Galactic and SMC reddening-to-dust ratio respectively and
extinction curves from Pei (1992).  These values are considerably higher
than the $E(B-K)$ observed for this source (see Figure \ref{norm_z}).
While there are a number of explanations for this discrepancy (e.g.
dustless gas near the QSO or AGN extinction curves that are significantly
different from Local Group galaxies, e.g. Gaskell et al. 2004)
a large reddening due to assoaciated absorption is very plausible.

\medskip

\begin{figure}
\centering
\includegraphics[scale=0.8]{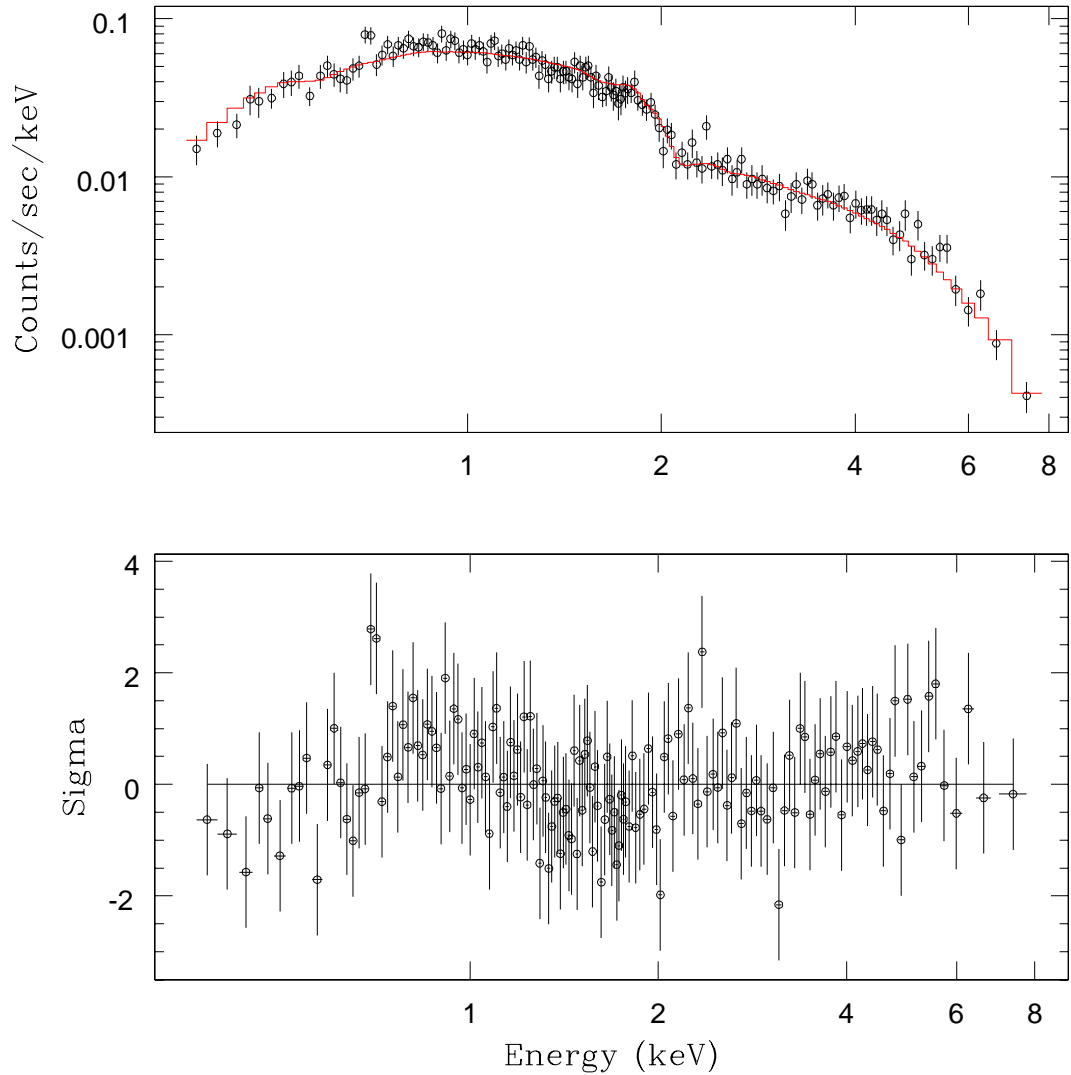}
\caption{\label{xray}Best-fit $\chi^{2}$ plots of the B0458$-$020 Chandra
observations. The fit and sigma ((data - model) / error) plots are
shown after convolving with the instrument response (top and bottom
panels, respectively). The data were fitted with a single power-law
model absorbed by a Galactic foreground \nhi\ column (thin
continous line in top panel).}
\end{figure}

We have retrieved archival Chandra ACIS data for B0458$-$020 (ObsID 2985)
with a total exposure of 77.76 ksec. The data were analysed in the
0.3-10.0 KeV energy range using CIAO 3.2.1 software and the latest
calibration files. B0458$-$020 suffers from relatively
high Galactic absorption,
corresponding to \nhi\ $= 7.5 \times 10^{20}$ cm$^{-2}$, which
hinders the detection of intrinsic absorption which is redshifted to
the soft (low energy) part of the spectrum. Indeed, a spectral fit
assuming a simple power-law spectrum with only Galactic absorption
yields a good fit to the
data, i.e.  $\chi^{2}_{\rm red} = 0.8$ for 155 degrees of
freedom. The best-fit photon index $\Gamma$ ($f_{\nu} \propto
\nu^{\Gamma+1}$ ergs s$^{-1}$ cm$^{-2}$) was found to be $1.55 \pm
0.02$ (see Figure \ref{xray}). The inclusion of absorption at either 
the QSO or the DLA
redshift gave no significant improvement to the fit: N(H) $< 2
\times 10^{21}$ cm$^{-2}$ with 99\% confidence (F-test statistics $F
~\sim 0.002$ for $\Delta \chi^{2} \sim 7.3$ and 1 additional
parameter). Using the Pei (1992) extinction laws this corresponds to
an upper limit $E(B-K) < 2.8$ and $< 0.7$, for a Galactic and SMC
reddening-to-dust ratio if the absorber is located in
the host galaxy.  Associated absorption could therefore be
the cause of the red color of this QSO, although better
X-ray data would be required to confirm this.  However, we note
that Carilli et al. (1998) found that 80\% of red QSOs in their
sample exhibited strong 21 cm absorption at the QSO redshift,
supporting the view that red colors may be often caused by intrinsic
absorption.
%If the absorber is located at the DLA redshift the
%corresponding color excess is $E(B-K) < 2.1$ and $< 0.6$, again for 
%Galactic and SMC reddening-to-dust ratio, respectively.   
%A Galactic reddening law is therefore sufficient
%to explain the observed $E(B-K)$ excess whether the absorber is
%located at the QSO rest frame or at the redshift of the DLA system.

\section{Discussion}

In a series of papers, Fall \& Pei (1989), Fall, Pei \& McMahon (1989)
and Pei, Fall \& Bechtold (1991) investigated whether the spectral
indices of QSOs with DLAs were systematically steeper that QSOs with
no intervening galaxy.  In particular, in the latter of these papers,
new data with higher spectrophotometric accuracy were obtained and
a sample of 66 QSOs analysed for reddening.  The distribution of
spectral indices in the DLA sample (20 QSOs) was found to differ
at the 99.999\% confidence level from the no-DLA (46 QSOs) sample.
The inferred dust:gas ratio was 5--20\% of that in the Milky Way.
Further support for observed reddening was reported by Outram
et al. (2001) who found redder B$-$R colors for QSOs with $0.5 < z < 1.5$
high equivalent width \ion{Mg}{2} absorbers.  The inferred color
excess was \ebmv\ $\sim$ 0.04.  More recently, Wild \& Hewett (2005)
have found a convincing reddening signature associated with low $z$ Ca~II
absorbers in the SDSS, although still at a modest level: \ebmv\ = 0.06.
Individual cases of DLA reddening
have been reported based either on differential reddening in
lensed systems (e.g. Zuo et al. 1997 for a DLA at $z_{\rm abs} = 1.4$) 
or the detection of the 2175
\AA\ graphite feature (Junkkarinen et al. 2004 for a DLA at 
$z_{\rm abs} = 0.5$).  However, some caution is required
when comparing inferences on dust extinction at different redshift 
ranges since the amount, source
and composition of dust may differ, for example as the dominant source
of dust shifts from Type II supernovae to lower mass stars (e.g. 
Hirashita et al. 2005).  Nonetheless, these works apparently provide
robust evidence for the systematic reddening of QSOs with intervening
absorption systems at both high and low redshifts.  

Recently, however, Murphy \& Liske (2004) re-visited the spectral 
fitting approach with a large statistical sample drawn from the SDSS.
In addition to its size ($\sim 1450$ QSOs, with 72 DLAs) this sample
has the advantage that its data acquisition is homogeneous
and the spectral indices are
fit over a wider wavelength range than was done in earlier work.
With these improvements, Murphy \& Liske (2004) found no steepening
of the spectral indices for QSOs with DLAs and determined a 
limit of \ebmv\ $<0.02$ (99\% confidence).  Although Murphy \& Liske adopt
a similar approach to Fall, Pei and collaborators in fitting
spectral indices, they do so with an order of magnitude larger
sample, wider wavelength coverage and uniform data acquisition.
Since QSOs have a wide range of intrinsic colors and spectral shapes,
using the broadest baseline possible is important for identifying
a systematic dust reddening.

For the first time, optical--IR colors can now contribute to the on-going
debate.  Although our sample is modest in size in comparison with the
SDSS (although comparable with the samples of Pei, Fall and collaborators), 
the broad wavelength coverage
provides significant leverage for detecting reddening.  For example,
whereas the \bmk\ color changes by 0.74 for an \ebmv=0.05 (SMC
extinction $z_{\rm em} = 3.0, z_{\rm abs} = 2.6$), the $B-R$ color 
changes only by 0.35.  Moreover, the sample of QSOs on which we conduct
this work is based on an optically complete survey of radio-loud
quasars.  Therefore, our sample of DLAs should not be subject to
selection effects associated with dust extinction.

Our results are
presented graphically in Figure \ref{norm_z} where we show the
normalized \bmk\ colors for QSOs with and without DLAs.   
The mean color of QSOs without DLAs is $-0.10$, compared
with +0.12 for the DLA sub-sample.  Both color distributions
have RMS scatters $\sim 0.5$.  Neither a student's T-test nor
a KS test find any significant difference in the \bmk\ colors
of the two sub-samples.
We determine a 3$\sigma$ limit of \ebmv\ $<0.04, 0.07$ for SMC and
Galactic extinction respectively based on the results of our KS
test simulations.  Although not quite as sensitive as
limits from the SDSS (Murphy \& Liske 2004), our sample is
optically complete and could potentially contain heavily
extinguished QSOs which would not have been fond in magnitude limited
samples.  However, the low \ebmv\ allowed by this work shows
that even in optically complete samples, the amount of reddening is
small.  Such low values of \ebmv\ are consistent with the small H$_2$
fractions observed in DLAs; whilst a handful of absorbers have
log $f(H_2) \sim -2$, most have log $f(H_2) < -5$ (Ledoux,
Petitjean \& Srianand 2003).  Indeed, Tumlinson et al. (2002) have shown
that low molecular fractions are common in Galactic and Magellanic
Cloud sightlines when \ebmv\ $\lesssim$ 0.08.

\begin{figure}
\centerline{\rotatebox{270}{\resizebox{13cm}{!}
{\includegraphics{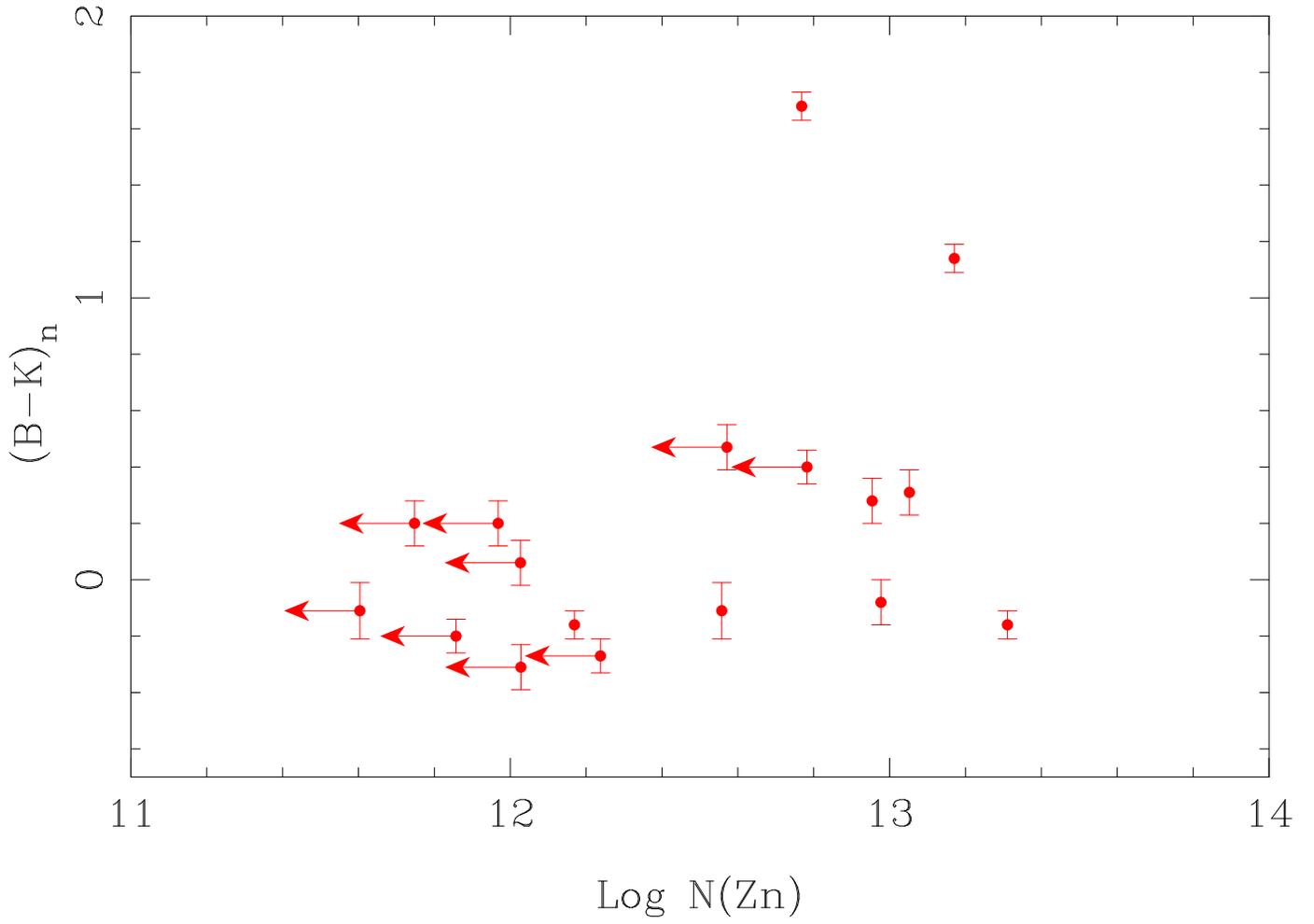}}}}
\caption{\label{bmk_metals2} Normalized \bmk\ color of CORALS DLAs as a
function of Zn column density. }
\end{figure}

\begin{figure}
\centerline{\rotatebox{0}{\resizebox{13cm}{!}
{\includegraphics{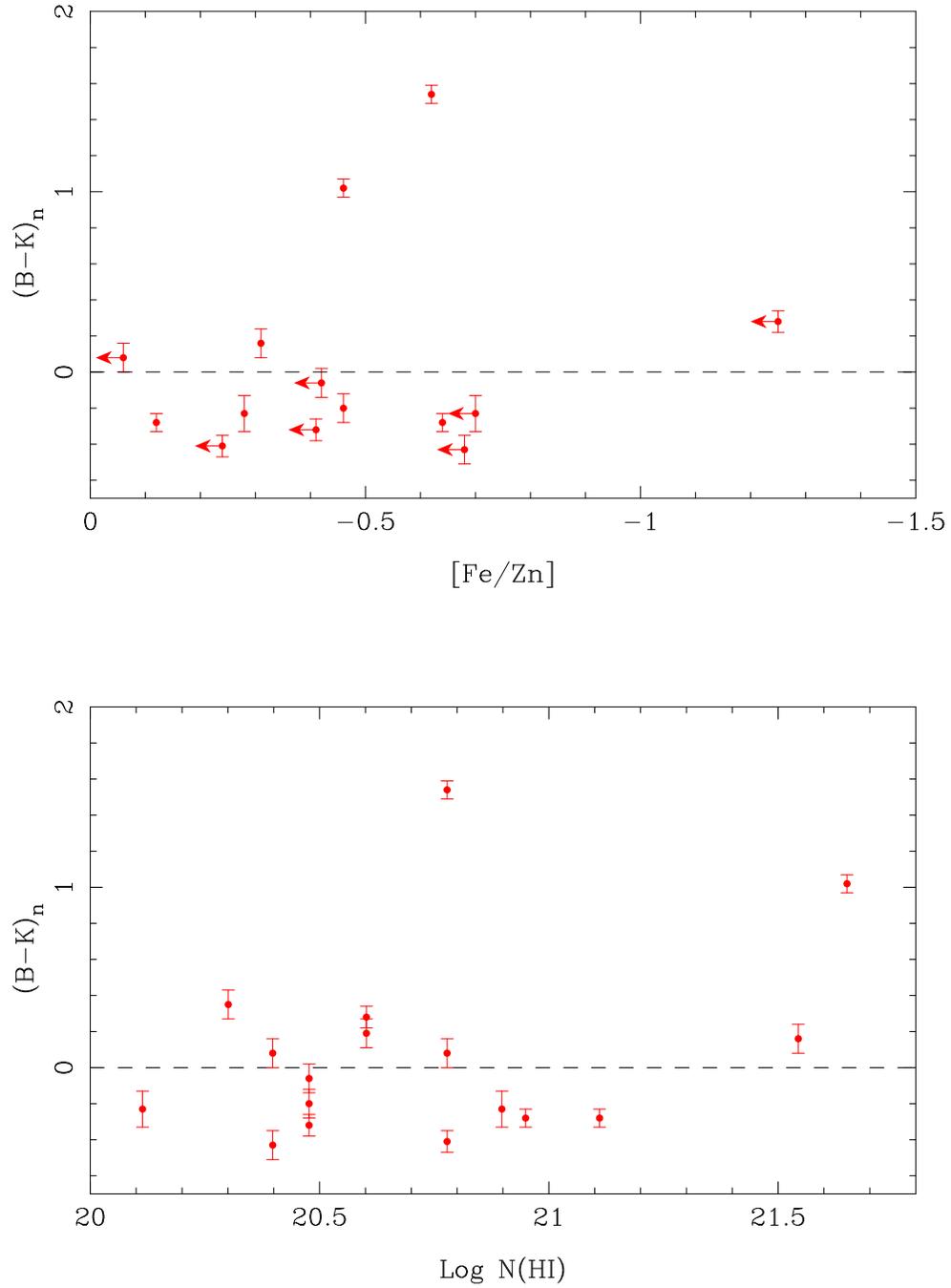}}}}
\caption{\label{bmk_metals}Top panel: Normalized \bmk\ color as a function
of depletion, as measured by [Fe/Zn].  Bottom panel: Normalized \bmk\ color 
as a function of neutral hydrogen column density, N(HI).}
\end{figure}

Although this work has not found any indication that QSOs are
significantly reddened by intervening DLAs, evidence that \textit{some} 
dust is present in DLAs
is well established from the abundances of refractory versus volatile 
elements such as Cr and Zn (e.g. Pettini et al. 1994, 1997).  
Akerman et al. (2005) have obtained
Zn and Fe abundances for many of the DLAs in our sample; we reproduce
these abundance results in Table \ref{norm_bmk_tab} adopting the solar
reference values from Lodders (2003).   Vladilo \&
P\'{e}roux (2005) have recently suggested that reddening will
increase linearly with N(Zn) and become especially severe when
log N(Zn) $>$ 13.2.  In Figure \ref{bmk_metals2} we plot the normalized
\bmk\ color versus N(Zn) using the column densities reported in
Akerman et al. (2005)\footnote{We make the usual assumption
that Zn~II is the dominant ionization stage of this element and
that the contribution from other ionzization stages is negligible.
Moreover, in Figures \ref{bmk_metals2} and \ref{bmk_metals} we include
2 DLAs within 3000 \kms\ of the QSO, plus one further absorber
whose \nhi\ is just below the canonical DLA limit.}.  
We detect no trend between normalized color
and log N(Zn), although we have only 3 DLAs with log N(Zn) $>$ 13.0.
Nonetheless, Akerman et al (2005) find that the mean weighted metallicity 
of CORALS DLAs is only marginally higher (although consistent
within the error bars) than previous samples.

In Figure \ref{bmk_metals} we investigate whether
there is any correlation between \nhi\ or depletion (as measured by
[Fe/Zn]) and reddening. 
The two reddest QSOs which also show DLA absorption have slightly larger 
than average ([Fe/Zn] = $-0.3$, Prochaska \& Wolfe 2002), but still modest
depletions.  The rest of the sample (including two systems with
comparable depletions to the reddest DLA-harboring QSOs) do
not show any correlation between depletion and reddening.  Similarly,
the N(HI) of one of the three very red QSOs in our sample has the DLA
with the highest \nhi\ amongst CORALS absorbers.  However, a second 
high \nhi\ DLA has apparently not
reddened its background QSO to a similar extent and the reddest QSO with
a DLA has only a modest \nhi.    Overall, there is
no obvious trend between \bmkn\ and \nhi.  On the other hand, Khare
et al. (2004) \textit{do} report tentative evidence for a correlation
between the inferred \ebmv\ (based on broad band SDSS colors) and
[Cr/Zn] (analogous to the [Fe/Zn] plotted in Figure \ref{bmk_metals}).
However, the Khare et al. result is based on the two systems
in their sample with [Cr/Zn] $<1$, and only one of the DLAs in our
sample may have such a high dust:gas ratio.  The normalized
\bmk\ color of this QSO is mildly redder than the median color of the
other QSOs, although this can certainly not be taken as evidence for
a correlation between depletion and reddening.  It would be of great interest
to further investigate the correlation between reddening and
dust:gas ratio in a larger sample of objects.

\begin{figure}
\centerline{\rotatebox{270}{\resizebox{13cm}{!}
{\includegraphics{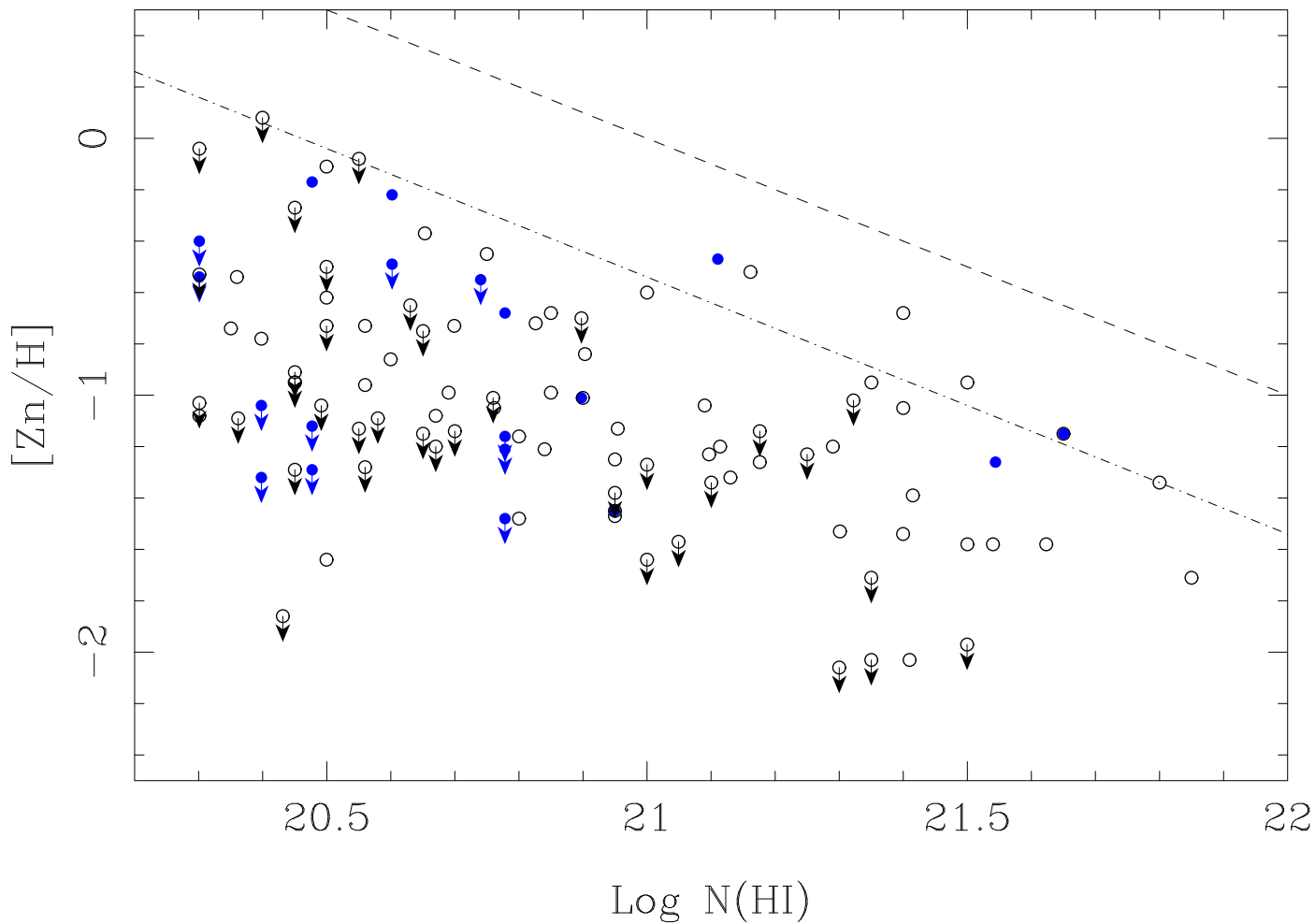}}}}
\caption{\label{znh_hi}Metallicity versus \nhi\ column density for DLAs
taken from the compilation of Kulkarni et al. (2005) and for the CORALS
DLAs (Akerman et al. 2005), shown by open and filled circles respectively.
The dashed line shows the proposed 'dust filter' of Prantzos \& Boissier
(2000): log \nhi\ + [Zn/H] $<$ 21, which corresponds to \ebmv\ $<$
0.17 for a Galactic gas-to-reddening law scaled by metallicity.  The
dot-dashed line shows where the cut-off would lie if \ebmv\ $<$ 0.05,
i.e. log \nhi\ + [Zn/H] $<$ 20.46. }
\end{figure}

Prantzos \& Boissier (2000) have suggested that dust bias is responsible
for the observed anti-correlation between [Zn/H] and \nhi, and propose
an empirical 'dust filter', i.e. that DLAs in optically selected samples 
all have log \nhi\ + [Zn/H] $<$ 21.  In Figure \ref{znh_hi} we plot \nhi\
versus [Zn/H] from the compilation of DLAs in Kulkarni et al. (2005);
%plus the abundances of CORALS DLAs published in Akerman et al. (2005).
the dust filter suggested by Prantzos \& Boissier (2000) is shown
by the dashed line.   We can combine the suggested cut-off of 
log \nhi\ + [Zn/H] $<$ 21 with a reddening-to-gas ratio
which scales with metallicity in order to calculate what \ebmv\
is implied.  For example,  it is 
straightforward to show that if the Galactic reddening-to-gas
ratio given in Eqn \ref{mw_eqn} is scaled by [Zn/H] and satisfies 
the criterion log \nhi\ + [Zn/H] $<$ 21, then \ebmv\ $<$ 0.17.   
This is the E(B-V) value required to ensure that no high N(HI), high
[Zn/H] systems are observed; \ebmv\ = 0.17 is indeed considerably
larger than the limits determined by this work and that of Murphy \&
Liske (2004).  We can alternatively calculate where the observed
cut-off in Figure \ref{znh_hi} would lie for a given assumed \ebmv.
For example, an \ebmv\ $<$ 0.05 implies log \nhi\ + [Zn/H] $<$ 
20.46 (again using Equation \ref{mw_eqn} scaled
by 10$^{\rm[Zn/H]}$); this cut-off is shown in Figure \ref{znh_hi} 
by a dot-dash line.  If future work determines an \ebmv\ $<<$ 0.05, 
it may (depending on the assumed reddening-to-gas ratio) become 
difficult to quantitatively reconcile the \nhi-[Zn/H] distribution 
with dust bias.

In Figure \ref{znh_hi} we also plot the \nhi\ and [Zn/H]
values for the CORALS DLAs, which should not be subject to dust bias.
The figure shows that CORALS DLAs are not distinguishable from other 
(literature)
DLAs in terms of their \nhi-[Zn/H] distribution.  Moreover, they do
not lie above the empirical cut-off which has been proposed to be
due to dust.    However, in the chemical evolution models
of Churches et al.(2004), $\lesssim$ 9\% of DLAs (depending on
the choice of spin parameter) are expected to be in this
'dust forbidden' region (A. Nelson, 2005, private communication).  
It therefore seems premature at this point to interpret the lack of 
high \nhi, metal-rich CORALS DLAs as evidence that the anti-correlation 
is not caused by dust.

If future work concludes that dust \textit{is} the reason for the 
anti-correlation
in Figure \ref{znh_hi}, then the locus of data points compared
with the theoretical lines and the lack of high CORALS values may
indicate that typical \ebmv\ values are simply intriniscally low
in most DLAs.  Indeed, Ellison, Kewley \& Mallen-Ornelas (2005)
found \ebmv\ $\lesssim$ 0.2 even in the central parts of absorber 
counterparts at $z < 0.5$.
Alternatively, the implied \ebmv\ values from this
and the work of Murphy \& Liske (2004) may be explained if the
dust extinction is relatively grey, as been proposed to explain the
\ebmv\ values observed towards some gamma-ray burst hosts 
(e.g. Vreeswijk et al. 2004).  

One of the greatest uncertainties in this work is the unknown nature
(i.e. composition, grain size and, therefore, extinction curve)
of dust in damped systems.  The main observational distinction between
SMC and Galactic extinction curves (with the LMC being somewhat intermediate
between the two, see Figure \ref{ext_curve}) is the presence of a 
'bump' at 2175 \AA, attributed to graphite.  A handful of detections
of this feature at cosmological redshifts exist associated with 3 Mg~II 
absorbers with $z_{\rm abs} \sim 1.5$ in the SDSS (Wang et al. 2004),
a high N(HI) DLA at  $z_{\rm abs} \sim 0.5$ (Junkkarinen et al. 2004) and
several cases of galaxy lenses (e.g. Falco et al. 1999; Toft et al. 2000;
Motta et al. 2002;  Wucknitz et al. 2003; Munoz et al. 2004).
However, gravitationally lensed systems are clearly special cases
since the impact parameter through the galaxy is necessarily small.
Malhotra (1997) have claimed to see the 2175 \AA\ feature in stacked 
(non-lensed) QSO spectra exhibiting \ion{Mg}{2} absorption, although
stacks of SDSS spectra have failed to reproduce this (M. Murphy, 2005,
private communication; Menard 2005; Khare et al. 2005; Wild \& Hewett 2005).  
Although further investigation is warranted,
this may be evidence that Galactic dust is not generally
applicable for absorbers towards unlensed QSOs.
Theoretical models of dust in high redshift galaxies enriched
by high mass Type II SN may be more relevant than local extinction
curves.  However, the extinction properties amongst these models
also vary tremendously depending on progenitor mass and mixing
(e.g. Hirashita et al. 2005).

\section{Conclusions}

In this work we have investigated whether dust in DLAs can cause
a significant, systematic reddening of background QSOs by obtaining
\bmk\ colors for an optically complete, radio-selected
quasar sample.  We determine normalized colors which account for
redshift dependence (analogous to a K-correction which accounts for \lya\
flux decrement and emission lines) and suppressed
B-band flux due to DLA absorption.
The mean normalized \bmk\ color of the DLA
sub-sample is 0.12$\pm0.56$, compared with $-0.10\pm0.47$ 
for the no-DLA sub-sample.  Both a student's T-test and KS test
indicate that there is no statistical difference between the
distribution of \bmk\ colors in the two sub-samples.
We place a limit on the amount of reddening that may be present,
by correcting the colors of QSOs with DLAs using a range of \ebmv\ values.
Adopting an SMC extinction curve, we place a 3$\sigma$
limit on the systematic reddening: \ebmv $<0.04$, assuming a 
fixed reddening for each DLA.  
%If we assume that the reddening depends on \nhi, we infer a
%reddening-to-gas ratio 30--50\% larger than the SMC (i.e. about
%15--20\% of the Galactic ratio).  
In general, there is no
correlation between depletion, N(Zn) or \nhi\ and reddening.
Finally, we have searched for archival X-ray data of the
three reddest QSOs in our sample in order to test whether their
extreme \bmkn\ colors maybe due to intrinsic (i.e. at $z \sim z_{\rm em}$)
absorption.  Of these three QSOs, two have data available, and
the spectrum of one (B0438$-$436) shows evidence for a high \nhi\
absorber associated with the QSO which may be the cause of the
very red color.

\medskip

These results support relatively low extinctions towards high
redshift DLAs, and agree with the \ebmv\ values inferred from
the optically selected SDSS work of Murphy \& Liske (2004).
Therefore, although a handful of individual absorbers may
cause more severe reddening (e.g. Junkkarinen et al. 2004),
these seem to be the exception rather than the rule.
A number of observational works have also inferred relatively small reddening
at lower redshifts (e.g. Khare et al. 2005; Menard 2005; Wild \& Hewett
2005).
Combined with previous work which has found that the number density
and gas content of intervening absorbers (Ellison et al. 2001, 2004)
and their metallicities (Akerman et al. 2005) are not largely different
from samples whose optical completeness exceeds $B \sim 19$, it seems
increasingly unlikely that dust bias has severely skewed our view of 
high redshift DLAs.  Put another way, the red(dened) QSO population
is unlikely to be linked to dust in high redshift DLAs.

\acknowledgements

The authors would like to thank the ever expert help of the La Silla
observatory staff, especially Valentin Ivanov.  SLE is grateful
to Jon Willis for providing advice regarding
the reduction of the SofI data presented here.  We are also
grateful to Paul Francis for making the data presented in FWW00
available in electronic format and for additional information
on that dataset.  SLE also acknowledges informative discussions
with Michael Murphy and is grateful for his feedback
on an earlier draft of this work.
PL acknowledges financial support by Fondecyt's grant 1040719.
This publication makes use of data products from the Two Micron All 
Sky Survey, which is a joint project of the University of Massachusetts 
and the Infrared Processing and Analysis Center/California Institute of 
Technology, funded by the National Aeronautics and Space Administration 
and the National Science Foundation.  This work also made use of the NASA 
Extragalactic Database (NED).

%% copy editing.  Individual instruments can be provided in parentheses,
%% after the keyword, but they will not be verified.

%Facilities: \facility{HST(STIS)}, \facility{CXO(ASIS)}.

% TABLE 1

\begin{deluxetable}{lccccc}
\tablewidth{6.0in}
\tablecaption{\label{obs_table}Target List and Observing Journal} 
\tablehead{
\colhead{QSO} & 
\colhead{$z_{\rm em}$} & 
\colhead{Observing} &
\colhead{Integration} & 
\colhead{Integration} &
\colhead{SuSI2}\\
\colhead{} &
\colhead{} &
\colhead{Run} &
\colhead{Time SofI (s)} &
\colhead{Time SuSI2 (s)}&
\colhead{Filter}
}
\startdata
 & & & & & \\
B0113$-$283 & 2.555 &  Oct 2003   & 1800 & 200 & B\\     
B0122$-$005 & 2.280 &  Oct 2003   & 300 & ... & ... \\
B0244$-$128 & 2.201 &  Oct 2003   & 300 & 40 & B\\   
B0256$-$393 & 3.449 &  Oct 2003   & 4320 & 360 & V\\  
B0325$-$222 & 2.220 &  Oct 2003   & 300 & 150 & B\\    
B0329$-$255 & 2.685 &  Oct 2003   & 200 & 40 & B\\      
B0335$-$122 & 3.442 &  Oct 2003   & 4500 & 360 & V\\   
B0347$-$211 & 2.944 &  Oct 2003   & ... & 800 & B\\   
B0405$-$331 & 2.570 &  Oct 2003   & 600 & 60 & B\\   
B0420$+$022 & 2.277 &  Oct 2003   & 800 & 100 & B\\   
B0422$-$389 & 2.346 &  Oct 2003   & 300 & 40 & B\\   
B0432$-$440 & 2.649 &  Oct 2003   & 800 & 100 & B\\     
B0434$-$188 & 2.702 &  Oct 2003   & 1800 & 200 & B\\    
B0438$-$436 & 2.863 &  Oct 2003   & 800 & 100 & B\\    
B0451$-$282 & 2.560 &  Oct 2003   & 300 & 60 & B\\   
B0458$-$020 & 2.286 &  Oct 2003   & 1200 & 200 & B\\    
B0528$-$250 & 2.765 &  Oct 2003   & 300 & 60 & B\\   
B0537$-$286 & 3.110 &  Mar 2004 & 1800 & 200 & B\\
          &       &     & & 200 & V \\ 
%B0537$-$286 & 3.110 &  Oct 2003   & & 200 & V\\   
B0610$-$436 & 3.461 &  Oct 2003   & 300 & 60 & V\\       
B0919$-$260 & 2.300 &  Mar 2004 & 300 & 60 & B\\    
B0933$-$333 & 2.906 &  Mar 2004 & 1800 & 200 & B\\
          &       &     & & 200 & V \\    
B1010$-$427 & 2.954 &  Mar 2004 & 800 & 20 & B\\   
          &       &      & & 20 & V \\    
B1055$-$301 & 2.523 &  Mar 2004 & 800 & 100 & B\\   
B1136$-$156 & 2.625 &  Mar 2004 & 800 & 200 & B\\   
B1147$-$192 & 2.489 &  Mar 2004 & 4500 & 360 & B\\    
B1149$-$084 & 2.370 &  Mar 2004 & 1800 & 200 & B\\   
B1228$-$113 & 3.528 &  Mar 2004 & 4920 & 800 & B\\  
          &       &     & & 800 & V \\    
B1228$-$310 & 2.276 &  Mar 2004 & 1800 & 200 & B\\   
B1230$-$101 & 2.394 &  Mar 2004 & 1800 & 200 & B\\    
B1256$-$243 & 2.263 &  Mar 2004 & 800 & 100 & B\\     
B1318$-$263 & 2.027 &  Mar 2004 & 4500 & 800 & B\\     
B1351$-$018 & 3.710 &  Mar 2004 & ...  & 800 & B\\ % Forgot to do K band!!   
B1354$-$107 & 3.006 &  Mar 2004 & 300 & 60 & B\\         
          &       &      & & 60 & V \\    
B1402$-$012 & 2.518 &  Mar 2004 & 300 & 20 & B\\     
B1418$-$064 & 3.689 &  Mar 2004 & 4500 & 360 & B\\    
          &       &      & & 360 & V \\  
B2311$-$373 & 2.476 &  Oct 2003   & 300 & 40 & B\\
\enddata
\end{deluxetable}

% TABLE 2 (Updated for Galactic extinction and correct 2MASS bootstraps)

\begin{deluxetable}{lcccccc}
\tablewidth{6.0in}
\tablecaption{\label{mag_table}CORALS QSO magnitudes} 
\tablehead{
\colhead{QSO} & 
\colhead{$z_{\rm em}$} & 
\colhead{$Bj_{UKST}$} &
\colhead{$B$} &
\colhead{$V$} &
\colhead{$K$} &
\colhead{Ref}
}
\startdata
B0017$-$307 &  2.666  &   20.0  &   ...             &    ...           &   ...             &    ...       \\
B0039$-$407 &  2.478  &   19.7  &   ...             &    ...           &   ...             &    ... 	 \\
B0104$-$275 &  2.492  &   19.3  &   ...             &    ...           &   ...             &    ...  	 \\
B0113$-$283 &  2.555  &   19.6  &   19.64$\pm$0.04  &    ...           &   17.32$\pm$0.15  &    This work \\
B0122$-$005 &  2.280  &   17.9  &   ...             &    ...           &   15.75$\pm$0.03  &    This work \\
B0244$-$128 &  2.201  &   18.4  &   18.21$\pm$0.04  &    ...           &   15.13$\pm$0.10  &    This work \\
B0256$-$393 &  3.449  &   20.6  &   ...             &   19.32$\pm$0.04 &   16.78$\pm$0.04  &    This work \\
B0325$-$222 &  2.220  &   19.0  &   19.37$\pm$0.04  &    ...           &   16.44$\pm$0.04  &    This work \\
B0329$-$255 &  2.685  &   17.6  &   18.11$\pm$0.04  &    ...           &   16.00$\pm$0.15  &    This work \\
B0335$-$122 &  3.442  &   20.6  &   ...             &   20.11$\pm$0.04 &   17.51$\pm$0.18  &    This work \\
B0347$-$211 &  2.944  &   20.5  &   20.89$\pm$0.04  &    ...           &   ...             &    This work \\
B0405$-$331 &  2.570  &   19.1  &   19.41$\pm$0.04  &    ...           &   16.28$\pm$0.04  &    This work \\
B0420$+$022 &  2.277  &   19.7  &   19.17$\pm$0.04  &    ...           &   16.37$\pm$0.04  &    This work \\
B0422$-$389 &  2.346  &   18.4  &   19.11$\pm$0.04  &    ...           &   16.60$\pm$0.13  &    This work \\
B0432$-$440 &  2.649  &   19.9  &   19.81$\pm$0.04  &    ...           &   17.37$\pm$0.04  &    This work \\ 
B0434$-$188 &  2.702  &   18.8  &   19.25$\pm$0.04  &    ...           &   16.24$\pm$0.07  &    This work \\
B0438$-$436 &  2.863  &   19.4  &   20.68$\pm$0.04  &    ...           &   16.09$\pm$0.03  &    This work \\
B0451$-$282 &  2.560  &   18.0  &   18.22$\pm$0.04  &    ...           &   15.35$\pm$0.03  &    This work \\
B0458$-$020 &  2.286  &   ...   &   19.33$\pm$0.04  &    ...           &   15.32$\pm$0.03  &    This work \\
B0528$-$250 &  2.765  &   18.1  &   18.16$\pm$0.04  &    ...           &   15.46$\pm$0.03  &    This work \\
B0537$-$286 &  3.110  &   20.0  &   19.84$\pm$0.07  &   19.06$\pm$0.04 &   16.16$\pm$0.03  &    This work \\
B0601$-$172 &  2.711  &   ...   &   ...             &    ...           &   ...             &    ...	 \\
B0610$-$436 &  3.461  &   18.5  &   ...             &   18.85$\pm$0.04 &   14.43$\pm$0.03  &    This work \\   
B0819$-$032 &  2.352  &   ...   &   ...             &    ...           &   ...             &    ...     	 \\
B0834$-$201 &  2.752  &   ...   &   ...             &    ...           &   ...             &    ...     	 \\
B0913$+$003 &  3.074  &   20.9  &   ...             &    ...           &   ...             &    ...	 \\
B0919$-$260 &  2.300  &   ...   &   18.05$\pm$0.07  &   ...            &   15.04$\pm$0.03  &    This work \\
B0933$-$333 &  2.906  &   ...   &   19.72$\pm$0.07  &   19.39$\pm$0.04 &   16.71$\pm$0.04  &    This work \\
B1010$-$427 &  2.954  &   ...   &   17.25$\pm$0.07  &   16.61$\pm$0.03 &   15.31$\pm$0.03  &    This work \\ 
B1055$-$301 &  2.523  &   19.3  &   19.36$\pm$0.07  &   ...            &   16.35$\pm$0.03  &    This work \\
B1136$-$156 &  2.625  &   19.1  &   19.31$\pm$0.07  &   ...            &   16.26$\pm$0.03  &    This work \\
B1147$-$192 &  2.489  &   20.3  &   20.38$\pm$0.07  &   ...            &   17.08$\pm$0.04  &    This work \\  
B1149$-$084 &  2.370  &   20.0  &   19.70$\pm$0.07  &   ...            &   16.52$\pm$0.04  &    This work \\
B1228$-$113 &  3.528  &   21.5  &   20.85$\pm$0.07  &   19.48$\pm$0.04 &   16.66$\pm$0.03  &    This work \\
B1228$-$310 &  2.276  &   19.8  &   19.36$\pm$0.07  &   ...            &   16.79$\pm$0.04  &    This work \\
B1230$-$101 &  2.394  &   19.6  &   19.73$\pm$0.07  &   ...            &   16.99$\pm$0.04  &    This work \\ 
B1251$-$407 &  4.464  &   ...   &   ...             &   ...            &   ...             &    ...       \\
B1256$-$243 &  2.263  &   19.4  &   19.45$\pm$0.07  &   ...            &   16.36$\pm$0.03  &    This work \\  
B1318$-$263 &  2.027  &   21.3  &   21.37$\pm$0.07  &   ...            &   17.57$\pm$0.03  &    This work \\ 
B1351$-$018 &  3.710  &   21.5  &   21.00$\pm$0.07  &   ...            &   ...             &    This work \\
B1354$-$107 &  3.006  &   18.4  &   19.00$\pm$0.07  &   18.15$\pm$0.04 &   15.70$\pm$0.03  &    This work \\  
B1402$-$012 &  2.518  &   18.0  &   18.50$\pm$0.07  &   ...            &   15.63$\pm$0.03  &    This work \\
B1406$-$267 &  2.430  &   ...   &   ...             &    ...           &   ...             &    ...       \\
B1418$-$064 &  3.689  &   20.5  &   20.42$\pm$0.07  &   18.97$\pm$0.04 &   16.58$\pm$0.03  &    This work \\
B1430$-$178 &  2.331  &   19.4  &   18.82$\pm$0.07  &   18.67$\pm$0.06 &   15.68$\pm$0.23  &    FWW00 	 \\
B1535$+$004 &  3.497  &   ...   &   ...             &    ...           &   ...             &    ...     	 \\
B1556$-$245 &  2.813  &   ...   &   18.46$\pm$0.06  &   18.69$\pm$0.06 &   16.53$\pm$0.31  &    FWW00 	 \\
B1635$-$035 &  2.871  &   ...   &   20.06$\pm$0.32  &   19.83$\pm$0.20 &   17.36$\pm$0.32  &    FWW00 	 \\
B1701$+$016 &  2.842  &   ...   &   ...             &    ...           &   ...             &     ...      \\
B1705$+$018 &  2.575  &   ...   &   18.30$\pm$0.07  &   18.07$\pm$0.06 &   15.85$\pm$0.10  &    FWW00 	 \\
B1937$-$101 &  3.780  &   ...   &   ...             &    ...           &   ...             &    ...       \\
B2000$-$330 &  3.780  &   ...   &   18.34$\pm$0.06  &   17.05$\pm$0.05 &   15.05$\pm$0.08  &    FWW00 	 \\
B2126$-$158 &  3.275  &   17.1  &   17.63$\pm$0.05  &   16.66$\pm$0.05 &   14.27$\pm$0.05  &    FWW00 	 \\
B2149$-$307 &  2.330  &   18.4  &   17.89$\pm$0.05  &   17.67$\pm$0.05 &   15.21$\pm$0.13  &    FWW00 	 \\
B2212$-$299 &  2.703  &   17.4  &   17.41$\pm$0.05  &   17.31$\pm$0.05 &   14.58$\pm$0.06  &    FWW00 	 \\
B2215$+$020 &  3.550  &   22.0  &   21.61$\pm$0.31  &   20.24$\pm$0.10 &   19.30$\pm$1.78  &    FWW00 	 \\
B2224$+$006 &  2.248  &   22.0  &   ...             &    ...           &   ...             &    ...       \\
B2245$-$059 &  3.295  &   19.7  &   ...             &    ...           &   ...             &    ...       \\
B2245$-$328 &  2.268  &   18.3  &   19.00$\pm$0.07  &   18.76$\pm$0.07 &   16.08$\pm$0.19  &    FWW00 	 \\
B2256$+$017 &  2.663  &   19.6  &   ...             &    ...           &   ...             &    ...       \\
B2311$-$373 &  2.476  &   18.4  &   18.99$\pm$0.04  &    ...           &   16.38$\pm$0.04  &    This work \\
B2314$-$340 &  3.100  &   18.3  &   ...             &    ...           &   ...             &    ...     	 \\
B2314$-$409 &  2.448  &   19.0  &   18.26$\pm$0.06  &   18.03$\pm$0.05 &   15.58$\pm$0.08  &    FWW00	 \\
B2315$-$172 &  2.462  &   21.0  &   ...             &    ...           &   ...             &    ...       \\
B2325$-$150 &  2.465  &   20.0  &   ...             &    ...           &   ...             &    ...       \\
B2351$-$154 &  2.665  &   18.7  &   18.82$\pm$0.07  &   18.60$\pm$0.06 &   16.37$\pm$0.26  &    FWW00  	 \\
\enddata
\end{deluxetable}

% TABLE 3 Normalized Colours and Absorber properties (abundances, redshifts, N(HI)) Corrected for
% Galactic extinction and DLA absorption

\begin{deluxetable}{lcccrrrc}
\tablewidth{7.0in}
\tablecaption{\label{norm_bmk_tab}Normalized Colors and Intervening Absorber Properties for CORALS QSOs with Near-Simultaneous \bmk\ Colors} 
\tablehead{
\colhead{QSO} & 
\colhead{$z_{\rm em}$} & 
\colhead{$z_{\rm abs}$} &
\colhead{N(HI)} &
\colhead{$(B-K)_n$} &
\colhead{[Zn/H]} &
\colhead{[Fe/H]} &
\colhead{Abundance Ref.}
}
\startdata
B0113$-$283 & 2.555 & ..... &  ..... & $-$0.49$\pm$0.16  & $ ..... $&$ .....$ & ..... \\
B0244$-$128 & 2.201 & ..... &  ..... &  0.17$\pm$0.11  & $ ..... $&$ .....$ & ..... \\
B0325$-$222 & 2.220 & ..... &  ..... &  0.02$\pm$0.06  & $ ..... $&$ .....$ & ..... \\
B0329$-$255 & 2.685 & ..... &  ..... & $-$0.68$\pm$0.16  & $ ..... $&$ .....$ & ..... \\
B0405$-$331 & 2.570 & 2.570 &  20.60 &  0.28$\pm$0.06  & $ <-0.49$&$ -1.74$ & Akerman et al. (2005) \\
B0420$+$022 & 2.277 & ..... &  ..... & $-$0.12$\pm$0.06  & $ ..... $&$ .....$ & ..... \\
B0422$-$389 & 2.346 & ..... &  ..... & $-$0.42$\pm$0.14  & $ ..... $&$ .....$ & ..... \\
B0432$-$440 & 2.649 & 2.297 &  20.78 & $-$0.41$\pm$0.06  & $ <-1.21$&$  -1.45 $ & Akerman et al. (2005) \\
B0434$-$188 & 2.702 & ..... &  ..... &  0.21$\pm$0.08  & $ ..... $&$ .....$ & ..... \\
B0438$-$436 & 2.863 & 2.347 &  20.78 &  1.54$\pm$0.05  & $ -0.68 $&$ -1.30$ & Akerman et al. (2005) \\
B0451$-$282 & 2.560 & ..... &  ..... &  0.07$\pm$0.05  & $ ..... $&$ .....$ & ..... \\
B0458$-$020 & 2.286 & 2.039 &  21.65 &  1.02$\pm$0.05  & $ -1.15 $&$ -1.61$ & Akerman et al. (2005) \\
B0528$-$250 & 2.765 & 2.141 &  20.75 & $-$0.28$\pm$0.05  & $ -1.45 $&$ -1.57$ & Akerman et al. (2005) \\
            &       & 2.811 &  21.20 &              & $ -0.47 $&$ -1.11$ & Akerman et al. (2005) \\
B0537$-$286 & 3.110 & 2.974 &  20.30 &  0.35$\pm$0.08  & $ <-0.40$&$ .....$ & Akerman et al. (2005) \\
B0919$-$260 & 2.300 & ..... &  ..... &  0.09$\pm$0.08  & $ ..... $&$ .....$ & ..... \\
B0933$-$333 & 2.906 & 2.682 &  20.48 & $-$0.06$\pm$0.08  & $ <-1.12$&$ -1.54$ & Akerman et al. (2005) \\
B1010$-$427 & 2.954 & ..... &  ..... & $-$1.14$\pm$0.08  & $ ..... $&$ .....$ & ..... \\
B1055$-$301 & 2.523 & 1.904 &  21.54 &  0.16$\pm$0.08  & $ -1.26 $&$ -1.57$ & Akerman et al. (2005) \\
B1136$-$156 & 2.625 & ..... &  ..... &  0.28$\pm$0.08  & $ ..... $&$ .....$ & ..... \\
B1147$-$192 & 2.489 & ..... &  ..... &  0.42$\pm$0.08  & $ ..... $&$ .....$ & ..... \\
B1149$-$084 & 2.370 & ..... &  ..... &  0.25$\pm$0.08  & $ ..... $&$ .....$ & ..... \\
B1228$-$113 & 3.528 & 2.193 &  20.60 &  0.19$\pm$0.08  & $ -0.22 $&$ .....$ & Akerman et al. (2005) \\
B1228$-$310 & 2.276 & ..... &  ..... & $-$0.35$\pm$0.08  & $ ..... $&$ .....$ & ..... \\
B1230$-$101 & 2.394 & 1.931 &  20.48 & $-$0.20$\pm$0.08  & $ -0.17 $&$   -0.63$ & Akerman et al. (2005) \\
B1256$-$243 & 2.263 & ..... &  ..... &  0.17$\pm$0.08  & $ ..... $&$ .....$ & ..... \\
B1318$-$263 & 2.027 & ..... &  ..... &  1.02$\pm$0.08  & $ ..... $&$ .....$ & ..... \\
B1354$-$107 & 3.006 & 2.501 &  20.40 &  0.08 $\pm$.08  & $ <-1.32$&$ -1.25$ & Akerman et al. (2005) \\
            &       & 2.966 &  20.78 &   & $ <-1.48$&$ -1.54$ & Akerman et al. (2005) \\
B1402$-$012 & 2.518 & ..... &  ..... &  0.02$\pm$0.08  & $ ..... $&$ .....$ & ..... \\
B1418$-$064 & 3.689 & 3.449 &  20.40 & $-$0.43$\pm$0.08  & $ <-1.04$&$ -1.72$ & Akerman et al. (2005) \\
B1430$-$178 & 2.331 & ..... &  ..... &  0.21$\pm$0.24  & $ ..... $&$ .....$ & ..... \\
B1705$+$018 & 2.575 & ..... &  ..... & $-$0.34$\pm$0.12  & $ ..... $&$ .....$ & ..... \\
B2000$-$330 & 3.780 & ..... &  ..... & $-$1.16$\pm$0.10  & $ ..... $&$ .....$ & ..... \\
B2126$-$158 & 3.275 & ..... &  ..... & $-$0.20$\pm$0.07  & $ ..... $&$ .....$ & ..... \\
B2149$-$307 & 2.330 & ..... &  ..... & $-$0.25$\pm$0.14  & $ ..... $&$ .....$ & ..... \\
B2212$-$299 & 2.703 & ..... &  ..... &  0.03$\pm$0.08  & $ ..... $&$ .....$ & ..... \\
B2245$-$328 & 2.268 & ..... &  ..... &  0.00$\pm$0.20  & $ ..... $&$ .....$ & ..... \\
B2311$-$373 & 2.476 & 2.182 &  20.48 & $-$0.32$\pm$0.06  & $ <-1.29$&$ -1.70$ & Akerman et al. (2005) \\
B2314$-$409 & 2.448 & 1.857 &  20.90 & $-$0.23$\pm$0.10  & $ -1.02 $&$ -1.33$ & Ellison \& Lopez (2001) \\
            &       & 1.875 &  20.10 &              & $ <-1.19$&$ -1.88$ & Ellison \& Lopez (2001) \\
B2351$-$154 & 2.665 & ..... &  ..... & $-$0.34$\pm$0.27  & $ ..... $&$ .....$ & ..... \\
\enddata
\end{deluxetable}

%
% FIGURES
%

\newpage

%\begin{figure}
%\centerline{\rotatebox{270}{\resizebox{13cm}{!}
%{\includegraphics{ks_test_dust_nhi.ps}}}}
%\caption{\label{ks_nhi} Ks probability that de-reddened QSOs with DLAs
%have the same \bmkn\ distribution as QSOs without DLAs.  Reddening
%dependent on \nhi\ is assumed according to Equation \ref{smc_eqn}
%multiplied by a coefficient, $C$.  That is, when $C=1$ the reddening-to-gas
%ratio is equivalent to that observed in the SMC.}
%\end{figure}

\end{document}